\documentclass[aps,reprint,twocolumn,floatfix,superscriptaddress]{revtex4-1}
\usepackage[utf8]{inputenc}
\usepackage{amsmath}
\usepackage{amssymb}
\usepackage{wasysym}
\usepackage{physics}                                
\usepackage{esint}
\usepackage{graphicx}
\usepackage{microtype}                              
\usepackage{siunitx}
\usepackage{dsfont}                                 
\usepackage{scalerel}
\usepackage{color}
\usepackage{hyperref}
\usepackage[export]{adjustbox}
\usepackage[caption=false]{subfig}
\usepackage{enumerate}
\usepackage{booktabs}
\usepackage{array}
\newcolumntype{P}[1]{>{\centering\arraybackslash}p{#1}}
\usepackage{setspace}                               

\usepackage{textcomp}

\usepackage[smallerops]{newtxmath}

\definecolor{darkerblue}{RGB}{33, 85, 168}

\newcommand{\mc}[1]{\mathcal{#1}}                   
\renewcommand{\v}[1]{\mathbf{#1}}                   
\def\Z2{$\mathbb{Z}_2$}                             
\def\U1{$\text{U}(1)$}                              

\DeclareMathOperator{\dist}{dist}
\DeclareMathOperator{\diag}{diag}

\hypersetup{
    colorlinks = true,
    linkcolor  = darkerblue,
    citecolor  = darkerblue,
    urlcolor   = darkerblue,
}
\begin{document}

\title{Coherent propagation of quasiparticles in topological spin liquids at
finite temperature}
\author{Oliver Hart}
\affiliation{T.C.M.~Group, Cavendish Laboratory,  JJ~Thomson Avenue, Cambridge CB3 0HE, United Kingdom}
\author{Yuan Wan}
\affiliation{Institute of Physics, Chinese Academy of Sciences, Beijing 100190, China}
\affiliation{Songshan Lake Materials Laboratory, Dongguan, Guangdong 523808, China}
\author{Claudio Castelnovo}
\affiliation{T.C.M.~Group, Cavendish Laboratory,  JJ~Thomson Avenue, Cambridge CB3 0HE, United Kingdom}
\date{September 2019}
%
%

\begin{abstract}
\setstretch{1.1}
The appearance of quasiparticle excitations with fractional statistics is a remarkable defining trait of topologically ordered systems. In this work, we investigate
the experimentally relevant finite temperature regime in which one species of quasiparticle acts as a stochastic background for another, more energetically costly, species that hops coherently across the lattice.
The nontrivial statistical angle between the two species leads to interference effects that we study using a combination of numerical and analytical tools.
In the limit of self-retracing paths, we are able to use a Bethe lattice approximation to construct exact analytical expressions for the time evolution of the site-resolved density profile of a spinon initially confined to a single site.
Our results help us to understand the temperature-dependent crossover from ballistic to quantum (sub-)diffusive behaviour as a consequence of destructive interference between lattice walks.
The subdiffusive behaviour is most pronounced in the case of semionic mutual statistics, and it may be ascribed to the localised nature of the effective tight-binding description, an effect that is not captured by the Bethe lattice mapping.
In addition to quantum spin liquids, our results are directly applicable to the dynamics of isolated holes in the large-$U$ limit of the Hubbard model, relevant to ultracold atomic experiments. A recent proposal to implement \Z2 topologically ordered Hamiltonians using quantum annealers provides a further exciting avenue to test our results.
\end{abstract}

\maketitle
%
%

\section{Introduction}

Quantum spin liquids (QSLs) are a fascinating phase of matter characterised
pragmatically by the absence of long-range order down to temperatures much smaller than the characteristic interaction energy in the system.
In magnetic materials, this behaviour is facilitated by strong quantum fluctuations within a macroscopically degenerate manifold of classical states, resulting from frustration between the constituent magnetic moments.
Such materials often host emergent gauge fields and point-like,
fractionalised quasiparticle excitations
with anyonic statistics~\cite{Wilczek2009}.
These exotic properties make QSLs interesting from a fundamental perspective, as well as having potential applications in the storage and processing of quantum information~\cite{Kitaev2003,Nayak2008}.

Experimentally, quantum spin liquid candidate materials (for a review, see Refs.~\cite{Balents2010,Knolle2018}) exhibit broad
continua in inelastic neutron scattering, indicative of fractionalisation of the emergent quasiparticles (spinons)~\cite{Han2012, Shen2016, Paddison2017}.
Although suggestive, this feature is not specific to QSLs,
and it is desirable to have more concrete experimental signatures of QSL
behaviour.
In this manuscript, we focus on nonzero temperatures, where a finite density
of excitations---obeying anyonic statistics---are thermally excited.
We take the stance that, rather than being a hindrance, finite temperature behaviour can in fact offer a number of signatures of fractionalisation and anyonic statistics, and thence of quantum spin liquid behaviour~\cite{Nasu2016, Yoshitake2016, Yoshitake2017-024438, Yoshitake2017-064433, Halasz2018}.

Specifically, we are interested in understanding the role of fractional statistics on the interplay between quasiparticle excitations, in the intermediate temperature range where one species of quasiparticle (visons) is thermally excited and acts as a stochastic background for another species (spinons), which are conversely sparse and hop coherently across the lattice. This is indeed a situation relevant to several realistic Hamiltonians for quantum spin liquids, where there is a large separation between the energy costs of different species of quasiparticle (one can think for example of quantum spin ice~\cite{Gingras2014}, Kitaev materials~\cite{Hermanns2018} and valence bond systems~\cite{Moessner2011}).
\begin{figure}[t]
    \centering
    \begin{minipage}{0.35\linewidth}
        \hspace{-0.25cm}\raisebox{2cm}{\subfloat{\includegraphics[valign=c,width=\linewidth]{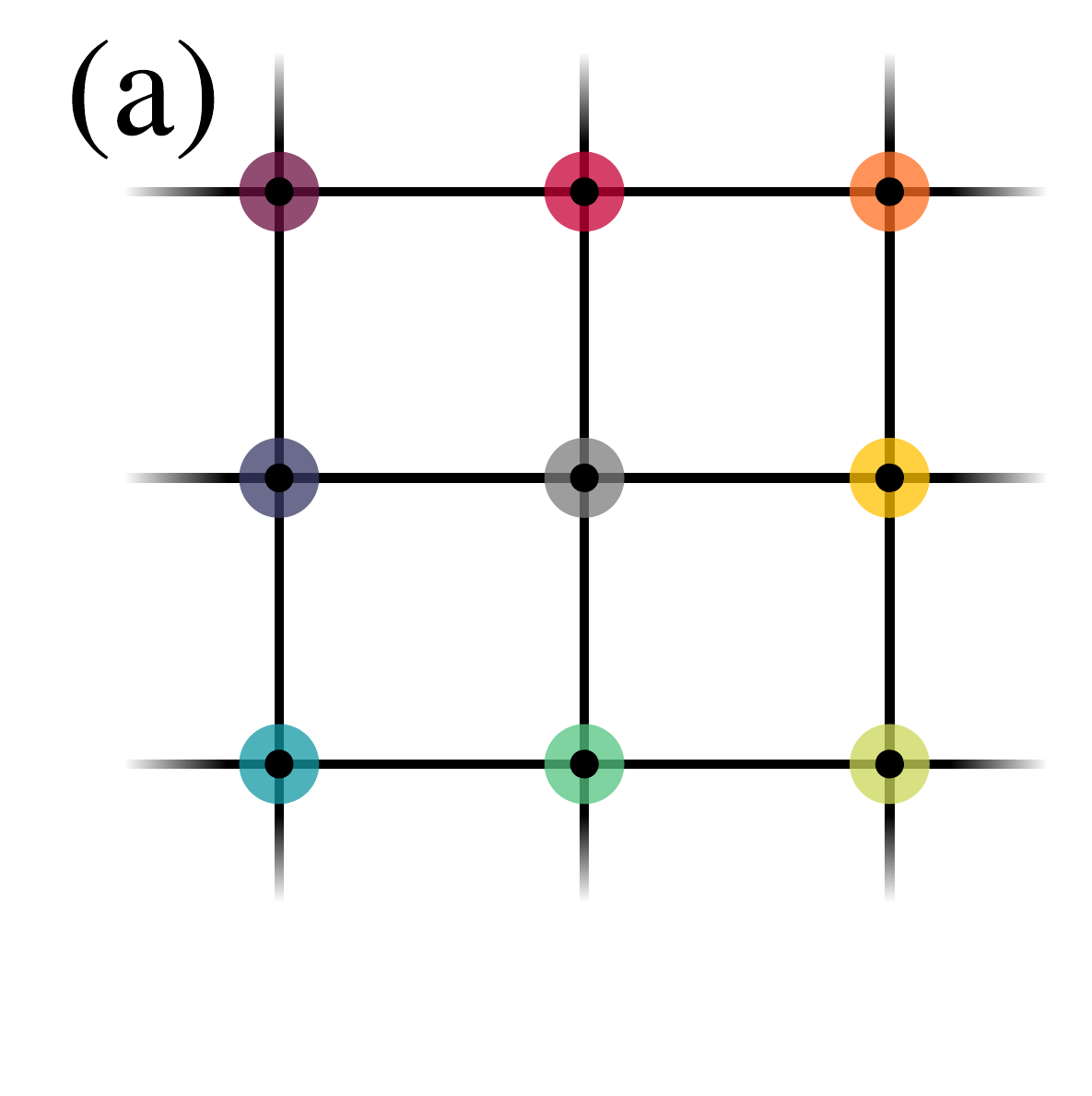}}}\\[-0.8cm]
        \subfloat{\includegraphics[valign=c,width=\linewidth]{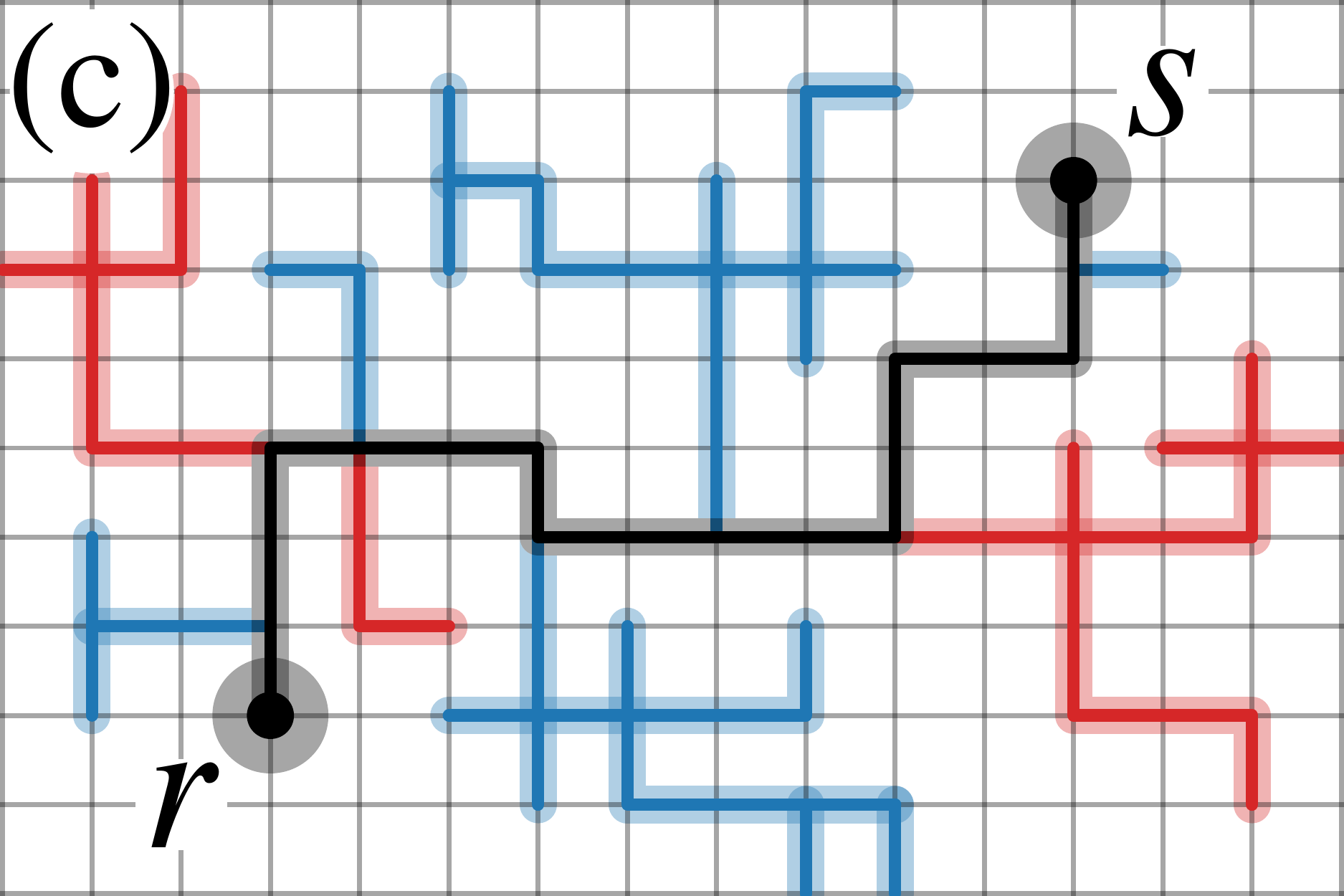}}
    \end{minipage}%
    \begin{minipage}{0.65\linewidth}
        \hspace{0.5cm}
        \subfloat{\includegraphics[valign=c,width=\linewidth]{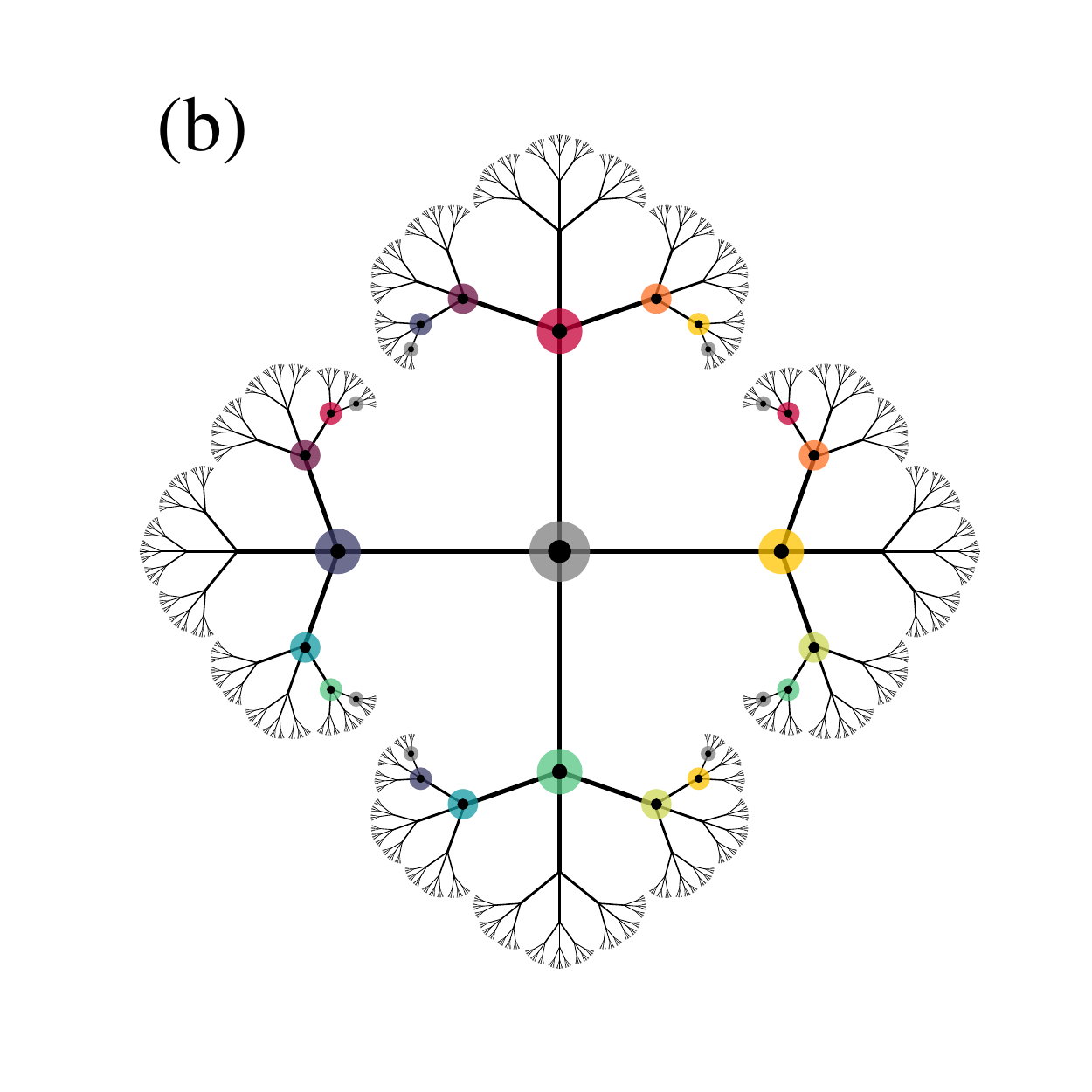}}
    \end{minipage}
    \caption{Mapping from the square lattice, (a), to the Bethe lattice with coordination number $z=4$, (b), used for the calculation of the spinon density profile.
    Each site on the square lattice is mapped onto multiple sites on the Bethe lattice, as indicated by the coloured circles. An example of a perfectly self-retracing round trip ($r \to s \to r$) on the square lattice is shown in (c). The nonreversing base path connecting $r$ and $s$ is represented by the thick black line, while the self-retracing excursions that decorate the base path on the outward (return) trip are shown as thick blue (red) lines. Such a walk encloses precisely zero area and contributes to the high-temperature expansion of the transition probability $P_{r \to s}(t)$.}
    \label{fig:bethe-square-mapping}
\end{figure}

We consider for simplicity the case of hardcore bosonic quasiparticles that have no mutual interactions but obey nontrivial mutual statistics, as is the case for instance in \Z2 models such as Kitaev's toric code~\cite{Kitaev2003}, or valence bond states that are represented by short-range dimer models on non-bipartite lattices.

In the toric code, the spinons and visons are mutually semionic, i.e., their statistical angle is $\theta = \pi$. Namely, the thermally excited visons act as a static, stochastic $\pi$-flux background for the spinons, which hop coherently on the sites of a square lattice.
Realistic Hamiltonians may require the inclusion of further effects, for instance due to interactions between the quasiparticles, or dynamical hopping terms for the visons, which we ignore in our discussion.
For comparison, we also briefly consider the case where the statistical angle is smaller, $\theta = 2\pi/n$, with $n=3,4,\ldots$, as well as the limiting case of continuous fluxes, corresponding, respectively, to $\mathbb{Z}_3,\mathbb{Z}_4$, and compact \U1 lattice gauge theory, relevant to doped Mott insulators, fractional quantum Hall effects, and vortex liquids~\cite{Ioffe1991,Pryor1992,Altshuler1992,Gavazzi1993}.

The central result of our work is an analytical solution within the self-retracing path approximation, in which the effect of the visons is to constrain the worldlines of the
spinons to live on a Bethe lattice.
Specifically, we derive analytical expressions for the spinon density profile as a function of space and time that agree quantitatively with the behaviour of the \U1 model,
capturing the short-time ballistic propagation, the crossover
to quantum diffusive behaviour and the non-Gaussian nature of the density profile.

Using numerical simulations, we highlight the importance of the localised
nature of the eigenstates of the underlying effective Hamiltonian
for the spinons.
In the \Z2 case, the spinons exhibit a crossover from ballistic to
\emph{sub}diffusive behaviour at some characteristic time that depends on
the density of visons.
For the \U1 model, the crossover is instead from ballistic to
quantum diffusive propagation, as predicted by the Bethe lattice mapping,
with only minor subdiffusive
corrections becoming evident at the largest times in our simulations.
We attribute this difference to the distinct localisation properties of
the two models; the intermediate-time behaviour of both models may be
ascribed to an increasing fraction of states reaching their localisation
length as time progresses.
Weaker localisation in the \U1 model implies that a
negligibly small fraction of states have reached their localisation
length over the timescales of our simulations, and we correspondingly
observe a regime in which the particle exhibits approximately diffusive
behaviour.

Our results connect directly to the propagation of holes in the large-$U$, large-spin limit of the Hubbard model~\cite{Brinkman1970,Carlstrom2016,Bohrdt2019}, and hence to the behaviour of related cold atomic systems~\cite{Kanasz-Nagy2017}.
The hole density profile may be probed directly in experiment owing to recent developments in quantum gas microscopy~\cite{Bakr2009, Sherson2010, Haller2015, Cheuk2015, Parsons2015, Edge2015}.
As the hole moves throughout the spin environment, it permutes the spins.
This ``interaction'' with the spin environment leads to dissipationless decoherence~\cite{Prokofev2000, Prokofev2006}---the propagation of the hole is significantly slowed despite there being no transfer of energy between the hole and the spins. Our system therefore provides a new setting in which to observe dissipationless decoherence. Additionally our Bethe lattice calculation extends the Brinkman--Rice argument in Ref.~\onlinecite{Carlstrom2016} in a way that agrees well with the numerical results therein to significantly larger distances and time scales.

Furthermore, it was recently proposed in Ref.~\onlinecite{Chamon2019} that the
toric code and similar $\mathbb{Z}_2$ spin liquid Hamiltonians may be
realised using quantum annealers (e.g., the commercially available
D-Wave machines~\cite{Boothby2018}) as emulators of topological states.
The implementation strategy works in the limit of a large star
constraint and a perturbative transverse field---namely, the
limit relevant to the physics explored in this manuscript.
The use of quantum annealers in this respect promises to provide new
avenues to benchmark and explore the type of phenomena that we have
uncovered, in a setting where Hamiltonian parameters, initial conditions
and time evolution can be explored to an exquisite level of accuracy
in system sizes that are beyond the reach of other
numerical methods.

The manuscript is structured as follows. In Sec.~\ref{sec:motivation} we introduce the Hamiltonian and its perturbative limit that will be the focus of this manuscript.
Specifically, we show that our Hamiltonian maps onto an effective bond-disordered tight-binding model, where temperature controls the strength of the disorder.
We further show that the motion of a single spinon through a sea of static visons may then be determined by enumerating lattice walks.
The generating functions pertaining to the self-retracing approximation of walks on a Bethe lattice are derived and presented in Sec.~\ref{sec:generating-functions}.
These functions are then used in the calculation of our exact results in Sec.~\ref{sec:results}, which we compare with numerical simulations in Sec.~\ref{sec:numerical-results}.
Finally, in Sec.~\ref{sec:conclusions} we draw our conclusions and
present an outlook for this work.
%
%

\section{\label{sec:motivation}
Model}

In this section we introduce the model and explain how various system observables may be calculated by enumerating certain classes of lattice walks.

For concreteness, we will focus our attention on a $\mathbb{Z}_2$ lattice gauge theory perturbed by a small transverse magnetic field $h$ in the $z$ direction, written in terms of spins $\boldsymbol{\sigma}_i$ which live on the bonds (labelled by the index $i$) of a square lattice with $N$ sites (labelled by the index $s$) wrapped around a cylinder
\begin{equation}
    H = - J \sum_s A_s - h \sum_i \sigma_i^z
    \, .
    \label{eqn:perturbed-hamiltonian}
\end{equation}
The star operator $A_s = \prod_{i \in s} \sigma_i^x$, and $i \in s$ denotes the spins on the four bonds surrounding the lattice site $s$. The coupling constant $J$~($\gg h$) is positive.
The model may however also be defined on other two-dimensional lattices, and we will later broaden the scope of this work to include the case where the spins live instead on a kagome lattice (see Appendix~\ref{sec:other-lattices}).
Treating the magnetic field $h$ perturbatively, we arrive at the following ring-exchange Hamiltonian in the ground state sector
\begin{equation}
    H_\text{eff}^{(0)} = -J \sum_s A_s - \frac{5}{16} \frac{h^4}{J^3} \sum_p B_p
    \, ,
    \label{eqn:effective-toric-code}
\end{equation}
up to a constant energy shift that arises due to virtual creation and annihilation of quasiparticle pairs. The plaquette operator $B_p = \prod_{i \in p} \sigma_i^z$,
where $i \in p$ denotes the four spins surrounding the plaquette $p$.
Therefore, the toric code Hamiltonian~\cite{Kitaev2003} is generated perturbatively and splits the macroscopic degeneracy of the ground state sector.

The ground state of the effective model~\eqref{eqn:effective-toric-code} is characterised by eigenvalues $+1$ for all (commuting) operators $A_s$ and $B_p$ (and has a topological degeneracy that is however immaterial for the purpose of the present work).
Excitations correspond to states in which plaquette operators $B_p$ and/or star operators $A_s$ have negative eigenvalues.
We will refer to the energetically costly star defects as spinons, and to the lower-energy plaquette defects as visons ($h^4 / J^3 \ll J$, since $J \gg h$, by construction).

Let us then consider the two spinon sector, relevant for the intermediate temperatures of interest, $T \ll h,J$.
The magnetic field $h$ makes the spinons dynamical
\begin{equation}
    H_\text{eff}^{(2)} = 4J - h \sum_{\langle s s' \rangle} \left( b_s^\dagger \sigma_{ss'}^z b_{s'}^{\phantom{\dagger}} + \text{h.c.} \right)
    \, ,
    \label{eqn:two-spinon-hamiltonian}
\end{equation}
where $\langle ss' \rangle$ denotes neighbouring sites on the square lattice, and $\boldsymbol{\sigma}_{ss'}$ is the spin on the bond connecting sites $s$ and $s'$.
Since the magnetic field is applied perpendicular to the $x$ axis, the vison configuration remains precisely static~\footnote{Introducing a nonzero projection of the magnetic field onto the $x$ axis, $h_x$, gives rise to vison dynamics on a timescale $h_x^{-1}$. However, provided that $h_x \ll h$, over the intermediate timescales of interest to the motion of the spinons, $h^{-1} \ll h_x^{-1}$, the visons may be treated as static and our results apply.}.
The operators $b^{\phantom{\dagger}}_s,b^\dagger_s$ are hardcore bosons representing the spinon excitations, which live on the sites of the lattice~\footnote{Note that the spins $\sigma_{ij}^x$ and the operators $b_i$ are not independent: $A_s = \exp(i\pi b^\dagger_s b_s)$.}. Crucially, each spinon hopping event is accompanied by a spin flip in the $\sigma^x$ basis. For a given vison configuration, and when considering gauge-invariant quantities, we can hence map~\eqref{eqn:two-spinon-hamiltonian} onto a nearest neighbour tight-binding model
\begin{equation}
    H_\text{eff}^{(2)}(\{ \phi_{ss'} \}) = 4J - h \sum_{\langle s s' \rangle} \left( b_{s}^\dagger e^{i \phi_{ss'}} b_{s'}^{\phantom{\dagger}} + \text{h.c.} \right)
    \, ,
    \label{eqn:two-spinon-hamiltonian-phases}
\end{equation}
where the Peierls phases $\phi_{ss'}$ are determined by the positions of the visons---each vison contributes a $\pi$-flux threading the plaquette on which it resides~\footnote{Within a cylindrical geometry, it is possible to choose a gauge in which the hopping amplitudes are real and uniform in one direction and acquire an appropriate minus sign in the orthogonal direction, according to the specific vison realisation.}.

We consider temperatures $T \ll h$, such that the quantum coherence of the spinons is not
significantly affected; note that this includes both the regime $T \ll h^4/J^3$ and $T > h^4/J^3$, so that the density of visons, $n_v$, spans the full range $0 \leq n_v < 1/2$.
We are further interested in the study of intermediate times where perturbation theory may be applied, namely, $t \ll J/h^2$. At times comparable to $J/h^2$ one must include next-nearest neighbour hopping processes in the effective Hamiltonian~\eqref{eqn:two-spinon-hamiltonian-phases}. Such processes remove the chiral symmetry of the effective Hamiltonian on bipartite lattices and thus may lead to modifications of the resulting long-time dynamics.

The motion of spinons in our model is equivalent to a quantum particle propagating through a background of randomly-placed $\mathbb{Z}_2$ fluxes. In this work, we also briefly consider for comparison the generic cases of other values of the threaded fluxes, $2\pi/n$, $n=3, 4, \ldots$, and, in particular, the limiting case of the so-called continuous flux model in which the fluxes $\phi$ threading the plaquettes are drawn from the uniform distribution over $\phi \in [0, 2\pi)$. Analogously to the random $\mathbb{Z}_2$ flux model, the model with random $2\pi/n$ fluxes arises from a $\mathbb{Z}_n$ lattice gauge theory in a similar finite temperature regime where the flux excitations are thermally populated whilst the elementary charge excitations remain coherent. Likewise, the continuous flux model describes the motion of a charged particle through an incoherent $\mathrm{U}(1)$ gauge field.
%
%

\subsection{\label{sec:motivation:greens-function}
Single-particle Green's function
           }

One quantity of interest is the single-spinon~\footnote{On account of their fractionalised nature, spinons are created in pairs. The single-particle properties are relevant if treating the two spinons as independent.} on-site Green's function for the effective Hamiltonian~\eqref{eqn:two-spinon-hamiltonian}, defined by
\begin{equation}
    G_{ii}(t) = \langle\langle b_i^{\phantom{\dagger}}(t) b_i^\dagger(0) \rangle\rangle
    \, ,
    \label{eqn:on-site-greens-function}
\end{equation}
where the double angled brackets $\langle\langle \,\cdots\, \rangle\rangle$ refer to both the quantum expectation value and thermal averaging over vison (flux) configurations. The on-site Green's function gives us access to the finite-temperature single-particle density of states $\rho(\omega)$ for spinons.
The form of $G_{ii}(\omega)$ within the self-retracing path
approximation is well known in the context of the $t$-$J_z$ model~\cite{Brinkman1970, Mohan1991, Starykh1996, Chernyshev1999}.
We include its derivation using the method of generating functions for the sake of completeness.
Our methodology may then be generalised to determine the off-diagonal elements of the Green's function, $G_{ij}(t)$, with $i \neq j$.
These quantities are not on their own gauge invariant and must be multiplied by the phases corresponding to a given lattice path connecting sites $i$ and $j$ in order to construct a gauge invariant quantity~\cite{Altshuler1992}.

Formally expanding $e^{-iHt}$ governing the time evolution in~\eqref{eqn:on-site-greens-function} as a power series in time, we are able to convert the problem into a summation over discrete lattice paths $\gamma$~\footnote{One may alternatively calculate $G_{ii}(\omega)$ as a power series in $1/\omega$ using similar methods, as in Ref.~\onlinecite{Brinkman1970}}, where the particle moves one lattice spacing per step. Integrating out the vison configurations, we arrive at
\begin{equation}
    G_{ii}(t) = \sum_{\ell=0}^{\infty} \frac{(iht)^{\ell}}{\ell! } \sum_{  \gamma \in \Gamma(\ell)  } e^{- A_{\gamma}/\xi^2(T)}
    \, ,
\end{equation}
where $\Gamma(\ell)$ is the set of all paths of length $\ell$ that begin and end at the site $i$, and $A_\gamma = \sum_p A_p(\gamma)$ is the ``area'' enclosed
by $\gamma$: each plaquette $p$ contributes an area $A_p(\gamma) =
[1 - (-1)^{w_p(\gamma)}]/2$ if it is encircled a total of $w_p(\gamma)$
times by the path $\gamma$.
In the continuous flux model, a given plaquette contributes only if $w_p(\gamma) = 0$.

The length scale $\xi(T)$ appearing in the exponential,
\begin{equation}
    \xi^2(T) = \frac{1}{-\ln\tanh[5 \beta h^4/(16 J^3)]}
    \label{eqn:vison-separation-lengthscale}
    \, ,
\end{equation}
corresponds approximately to the average distance between visons,
$\sim \! n_v^{-1/2}$, in the dilute-vison limit, where $n_v \!\sim\! e^{-10\beta h^4/16J^{3}}$.
Note that $\xi \to 0^+$ for high temperatures, corresponding to the vison-dense limit, $n_v \to 1/2$.
At \emph{any} nonzero temperature, paths that enclose a large area with respect to $\xi^2$ are exponentially suppressed~\footnote{This may be viewed as an analogue of the high-temperature area law for loop correlation functions $\ev*{\prod_{\ell \in C} \sigma^z(\ell)} \!\sim\! \exp(-A)$ in lattice gauge theory~\cite{Kogut1979}.}, a manifestation of the Aharonov--Bohm effect. Recall that the zero-temperature ($n_v=0$) dynamics of the spinon is equivalent to a free quantum particle at all times $t$.
The limits of infinite time and zero temperature therefore do not commute.

From the ordinary generating function $g(x; a) = \sum_{n, m} g_{nm} x^n a^m$ for walks $\gamma \in \Gamma(\ell)$, where the generating variables $x$ and $a$ are associated with path length $n$ and area enclosed $m$, respectively, one can observe that $G_{ii}(t)$ is equal to the corresponding exponential generating function $\tilde{g}(x; a) \equiv \sum_{n, m} g_{nm} x^n a^m / n!$ via
\begin{equation}
    G_{ii}(t) = \tilde{g}(iht; e^{-1/\xi^2})
    \, .
\end{equation}
The effect of changing temperature is to alter the relative weight of the different lattice walks, classified according to the area that they enclose.
At zero temperature, all paths of a given length contribute with equal weight, while at infinite temperature only those paths that enclose precisely zero area contribute.
We expect that the latter result describes the limiting behaviour for high temperatures, $T > h^4/J^3$.
%
%

\subsection{Density evolution}

Motivated by the study of finite temperature dynamical spin--spin correlators, we would like to quantify the propagation of a pair of spinons after being created locally on adjacent sites.
As a first approximation, we solve the single-particle problem, which we are able to treat analytically.
In particular, we calculate (i) the site-resolved density profile for short times, $ht = O(1)$, accessible for instance in ultracold atomic experiments, and (ii) the asymptotic moments of the density distribution, $\langle \v{r}^{2k}(t) \rangle$, which characterise the behaviour of the spinon profile over a large range of intermediate timescales.

The (gauge invariant) probability for the spinon to move from site $0$ to site $s$ in a time $t$ in the presence of a given vison (flux) configuration $\{ \phi_p \}$ is given by
\begin{align}
    P_{s}(\{ \phi_p \}; t) &= \mel*{\{\phi_p\}}{b_0^{\phantom{\dagger}}(0) b_s^\dagger(t) b_s^{\phantom{\dagger}}(t) b_0^\dagger(0)}{\{\phi_p\}} \\
    &= \left\lvert \mel*{\{\phi_p\}}{b_s e^{-iHt} b_0^\dagger}{\{\phi_p\}} \right\rvert^2
    \, .
\end{align}
In a similar manner to $G_{ii}(t)$, we are able to write the transfer probability $P_s$ in terms of summation over outward ($\gamma$) and return ($\gamma^\prime$) lattice paths.
After integrating over the possible flux configurations $\{ \phi_p \}$ with the appropriate Boltzmann weight, the probability reads
\begin{equation}
    P_{s}(t) = \sum_{\ell, \ell^\prime = 0}^{\infty} (-1)^\ell \frac{(iht)^{\ell + \ell^\prime}}{\ell! \ell^\prime !} \sum_{ \substack{ \gamma \in \Gamma_s(\ell) \\ \gamma^\prime \in \Gamma_s(\ell^\prime) } } e^{- A_{\gamma \cup \gamma^\prime}/\xi^2}
    \, ,
\end{equation}
where $\Gamma_s(\ell)$ is the set of all paths of length $\ell$ that connect the sites $0$ and $s$, and $A_{\gamma \cup \gamma^\prime}$ is the area enclosed by the closed path $\gamma \cup \gamma^\prime$.
Knowledge of $P_s(t)$ for all sites $s$ gives us complete information about the density distribution $\rho(\v{r}, t)$ as a function of time.
%
%

\subsection{Interpretation}

We have shown that in both instances the problem of determining single-spinon motion in a sea of thermally excited visons may be mapped onto the combinatorial problem of enumerating discrete lattice walks.
At precisely zero temperature, the system is free of vison excitations, $\xi = \infty$, and all paths of a given length contribute with equal weight~\footnote{Note that the limits of infinite time and zero temperature do not commute. We are here referring to taking $T \to 0$ before $t \to \infty$.}.
In this limit, the effective Hamiltonian is simply a two-dimensional tight-binding model with nearest neighbour hopping, and the spinon propagates ballistically.
Conversely, at temperatures which are high with respect to the energy cost for vison creation, $T > h^4/J^3$, the hopping amplitudes are maximally binarily disordered, and only walks that enclose exactly ``zero area'' (as defined previously) contribute.
(Note that the notion of zero enclosed area trivially extends to the case of fluxes of magnitude $2 \pi m / n$, with $m, n \in \mathbb{N}$, threading the plaquettes.)
We focus primarily on this high-temperature limit in order to contrast with the known behaviour at $T=0$.

Exact enumeration of all such zero-area paths on a generic lattice with coordination number $z$ is a very tall order.
In order to tackle this problem, we discuss a limit where the problem becomes analytically tractable.
Namely, we consider perfectly self-retracing paths~\cite{Brinkman1970}, which necessarily satisfy $A_\gamma = 0$.
As we shall see, this is a particularly relevant subset of walks the smaller the threaded fluxes are, and particularly for the continuous \U1 flux case.
For comparison, in Sec.~\ref{sec:numerical-results}, we also solve numerically for the time evolution generated by effective
Hamiltonians of the form~\eqref{eqn:two-spinon-hamiltonian-phases} using a high order Suzuki--Trotter decomposition~\cite{DeRaedt1987}.
We will also discuss the lower temperature regime $T \lesssim h^4/J^3$ in the context of these simulations.
%
%

\section{Self-retracing paths: Generating functions}
\label{sec:generating-functions}

\begin{figure}
    \centering
    \subfloat[\label{fig:paths:a}]{%
        \includegraphics[width=.3\linewidth]{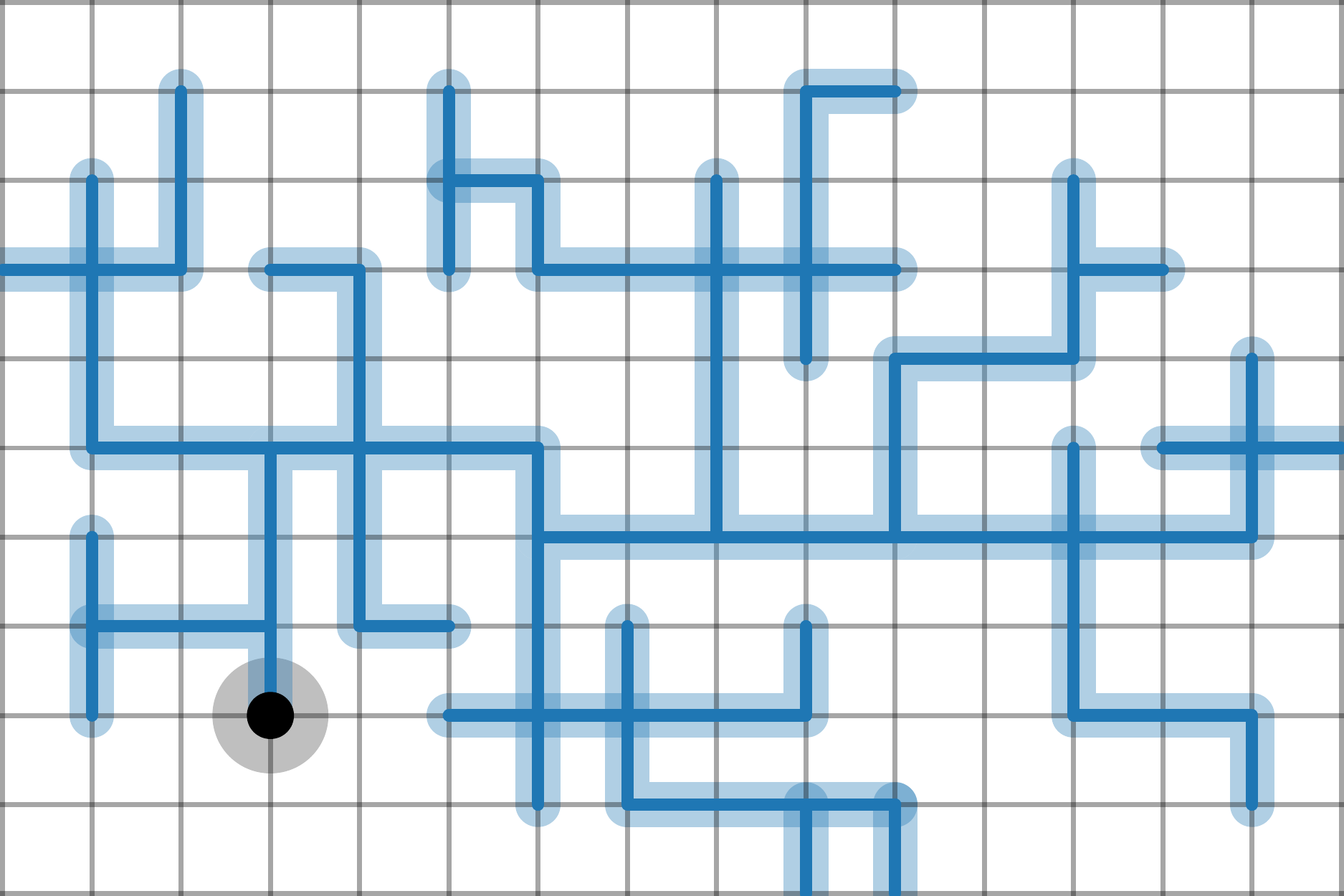}%
    }\hspace{0.1cm}
    \subfloat[\label{fig:paths:b}]{%
        \includegraphics[width=.3\linewidth]{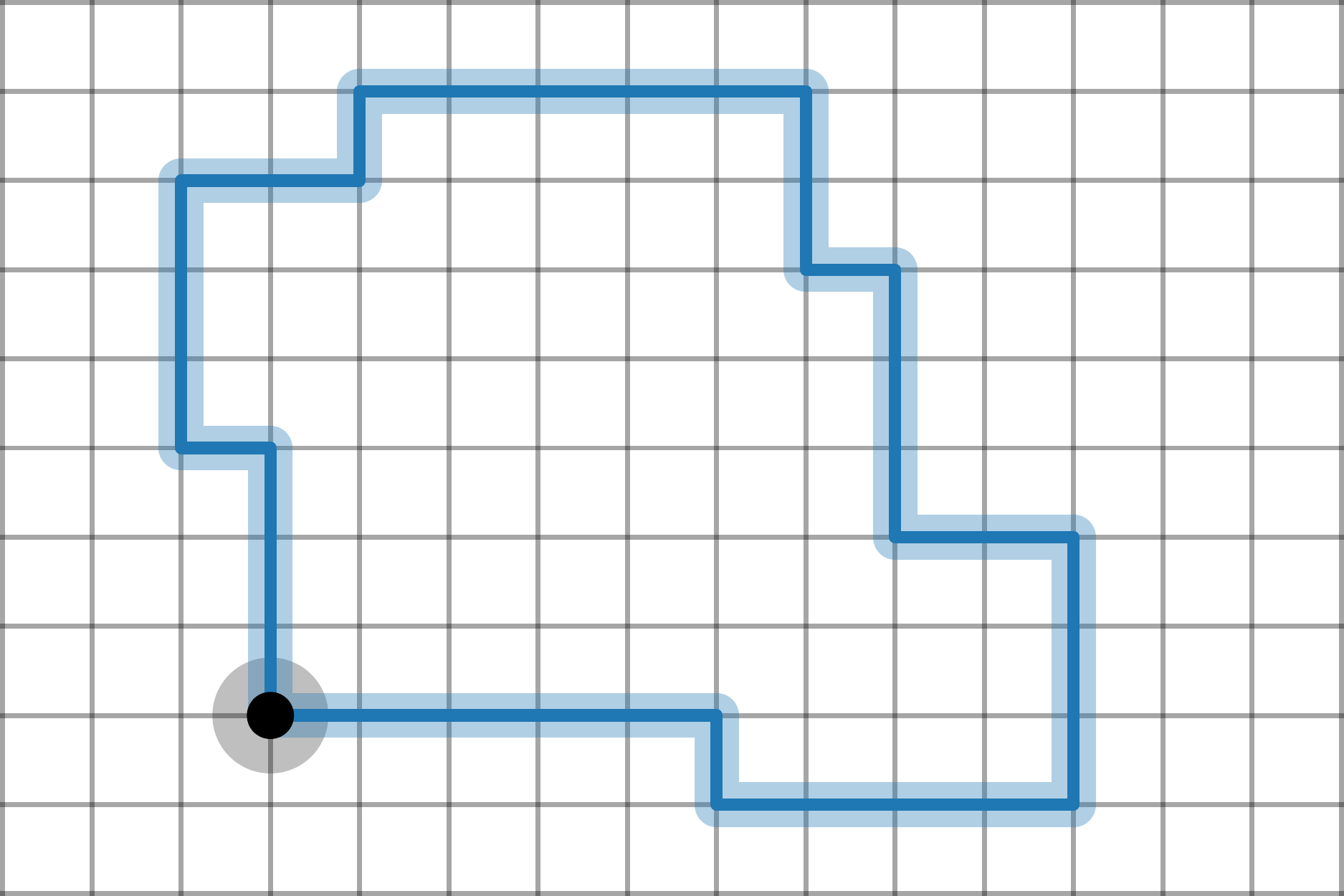}%
    }\hspace{0.1cm}
    \subfloat[\label{fig:paths:c}]{%
        \includegraphics[width=.3\linewidth]{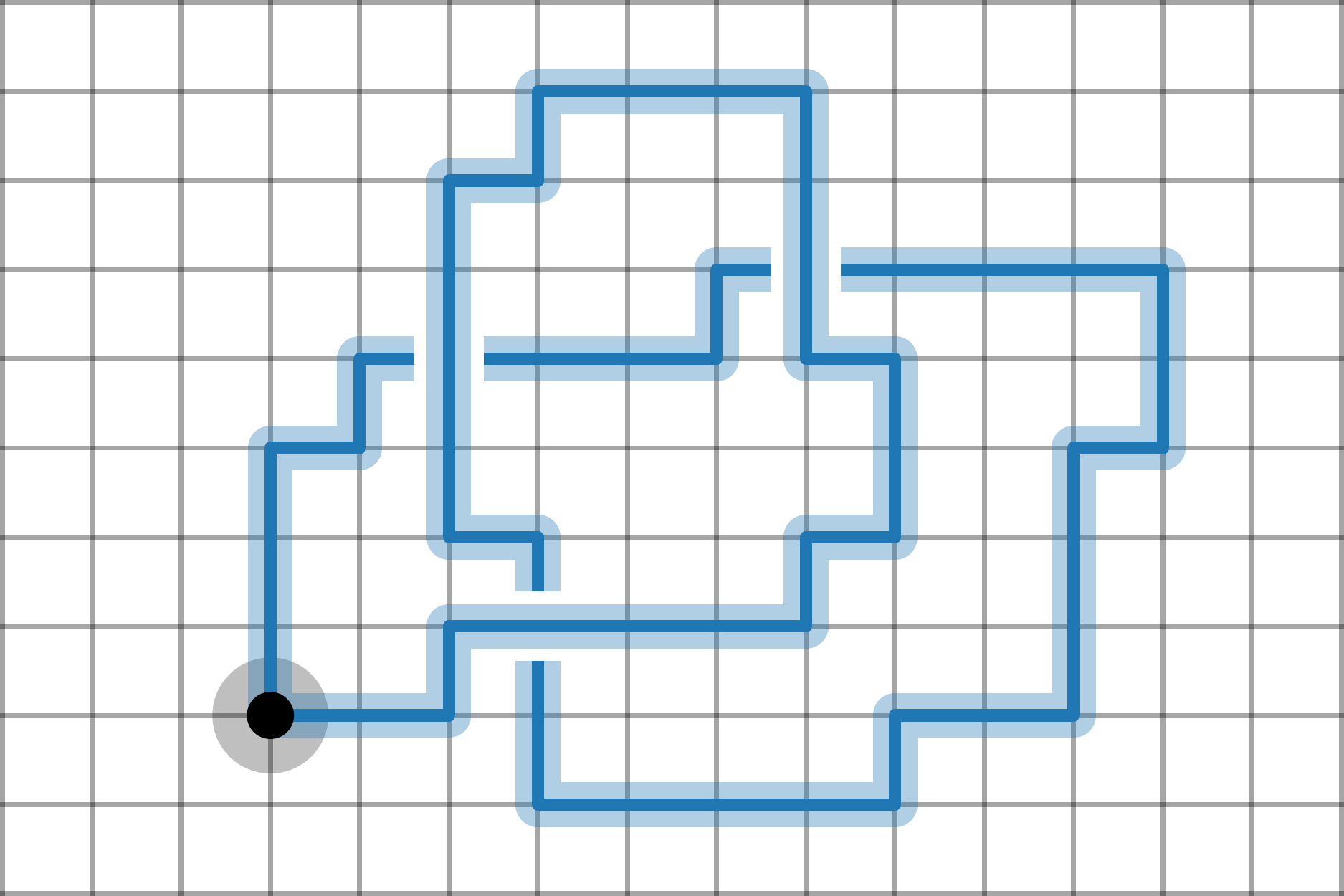}%
    }%
    \caption{Three examples of closed lattice walks, $\gamma$, which begin and end on the black circle.
    A perfectly self-retracing path---the only type of path included in the Bethe lattice mapping---is shown in (a).
    In (b) and (c) the walk includes closed cycles, which have the potential to be non-self-retracing.
    In general, a walk $\gamma$ contributes to the lattice path expansion at high temperatures if $\langle \exp \small(i \sum_{\langle ij \rangle \in \gamma} \phi_{ij}\small) \rangle_{\{\phi_{ij}\}} = 1$.
    In the case of $\pi$-fluxes threading the plaquettes, a walk in which the loop (b) is traversed an even number of times in the \emph{same} direction leads to a nonzero contribution, ${\langle e^{2ni\phi} \rangle}_{\mathbb{Z}_2} = 1$.
    Such a path with winding number $w=2n$ ($n \in \mathbb{Z}_{\neq 0}$) is \emph{not} self-retracing, and so is not captured by the Bethe lattice mapping.
    For continuous fluxes, however, such non-self-retracing paths of the form (b) with nonzero winding number $w=2n$ are not present in the lattice path expansion after averaging over flux configurations, since ${\langle e^{2ni\phi} \rangle}_{\text{U(1)}} = 0$.
    Paths that self-intersect multiple times, as in (c), can be traversed in multiple ways in the reverse direction, only one of which is self-retracing.
    All other paths are not accounted for by the Bethe lattice mapping, whereas they do however contribute to the continuous flux case.
    For these reasons, we expect the Bethe lattice mapping to better approximate the continuous flux model where a significantly larger fraction of permitted lattice walks are correctly enumerated.
    }
    \label{fig:paths}
\end{figure}
A perfectly self-retracing path corresponds to a lattice walk with no closed cycles in which every link on the `outwards' path is retraced in the opposite direction on the `return' path.
More precisely, there is a one-to-one mapping between self-retracing paths on a lattice, $\mc{L}_z$, with coordination number $z$, and closed walks on a Bethe lattice, $\mc{B}_z$, with branching ratio $z-1$, as shown for the case $z=4$ in Fig.~\ref{fig:bethe-square-mapping}.

Fractal lattices in general are a useful tool for obtaining exact solutions, and have recently been used in a similar context to calculate the spectrum of itinerant excitations in quantum spin ice at zero temperature~\cite{Petrova2015,Udagawa2018}, where gauge field effects lead to a configuration space which is well-approximated by the Husimi cactus graph.
Examples of perfectly self-retracing walks, and classes of walks which are \emph{not} captured by the self-retracing path approximation are shown in Fig.~\ref{fig:paths}.
The latter are analogous in spirit to the Trugman path~\cite{Trugman1988} in the context of single hole propagation in the $t$-$J$ model.
Since the paths that are not accounted for only become relevant at long times, one may expect that this approximation works well for the dynamics of the system at the intermediate timescales of interest, at least for the continuous flux model.

In this section we will derive the generating functions for walks on a Bethe lattice with branching ratio $z-1$, which are necessary to describe analytically the form of the density profile in the limit of high temperature (high flux density).
%
%

\subsection{Closed walks}
\label{sec:closed-walks}

We first consider the ordinary generating function $T_0^{(z)}(x) = \sum_n t_n^{(z)} x^n$ for closed walks on a Bethe lattice $\mc{B}_z$ with branching ratio $z-1$, where, by definition, $t_n^{(z)}$ is the number of closed walks that begin and end at the same site, which may be used to define the root node (or origin) of the Bethe lattice.
This generating function is directly related to the single-particle density of states.
Note that the lack of closed cycles implies that all closed walks on $\mc{B}_z$ are necessarily self-retracing, and further permits the following decomposition of the generating function
\begin{equation}
    T_0^{(z)}(x) = 1 + zx^2 T_0^{(z)}(x) T_1^{(z)}(x)
    \, ,
\end{equation}
where $T_k^{(z)}(x)$ is the generating function for walks beginning and ending at a depth of $k$ on the lattice (always remaining at a depth $\geq k$). This is because any (self-retracing) path can be decomposed as
\begin{enumerate}[(i)]
    \item the trivial walk,
    \item \begin{enumerate}[(a)]
        \item hopping to one of the $z$ nearest neighbours,
        \item performing a self-retracing walk that begins and ends at depth $k=1$,
        \item hopping back to the origin,
        \item performing a self-retracing walk that begins and ends at the origin.
    \end{enumerate}
\end{enumerate}
A similar argument can be made for all subsequent depths with $k \geq 1$, such that the generating functions decompose as
\begin{equation}
    T_{k}^{(z)}(x) = 1 + (z-1)x^2 T_k^{(z)}(x)T_{k+1}^{(z)}(x)
    \, .
\end{equation}
One can therefore express the original generating function $T_0^{(z)}$, for paths beginning and ending at the origin, as an infinite continued fraction
\begin{equation}
    T_0^{(z)}(x) = \cfrac{1}{1-\cfrac{zx^2}{1-\cfrac{(z-1)x^2}{1-\cfrac{(z-1)x^2}{1-\ddots}}}}
        \, .
\end{equation}
The self-similar nature of $\mc{B}_z$ implies that $T_k^{(z)}(x) = T_{k+1}^{(z)}(x)$ for $k \geq 1$ (on an infinite lattice), and the continued fraction can be written in closed form (choosing the sign in front of the square root such that $T_0^{(z)}(x) \to 1$ as $x \to 0$)
\begin{equation}
    T_0^{(z)}(x) = \frac{2(z-1)}{z-2+z\sqrt{1-4(z-1)x^2}}
    \, ,
    \label{eqn:ordinary-generating-function}
\end{equation}
consistent with, e.g., the results of Ref.~\onlinecite{Wanless2010}. For $z=4$ (corresponding to the square lattice at high temperatures), this expression evaluates to
\begin{align}
    T^{(4)}_0(x) &= \frac{3}{1+2\sqrt{1-12x^2}} = 1 + 4x^2 + 28x^4 + \ldots \\
    &= \includegraphics[height=4pt, valign=c]{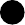} + (4 \, \includegraphics[height=4pt, valign=c]{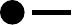})x^2 + (16 \, \includegraphics[height=4pt, valign=c]{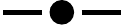} + 12 \, \includegraphics[height=4pt, valign=c]{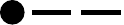})x^4 + \ldots
\end{align}
where the diagrams denote the types of self-retracing walk that contribute at each order.
%
%

\subsection{Open walks}

\begin{figure}[t]
    \centering
    \includegraphics[width=\linewidth]{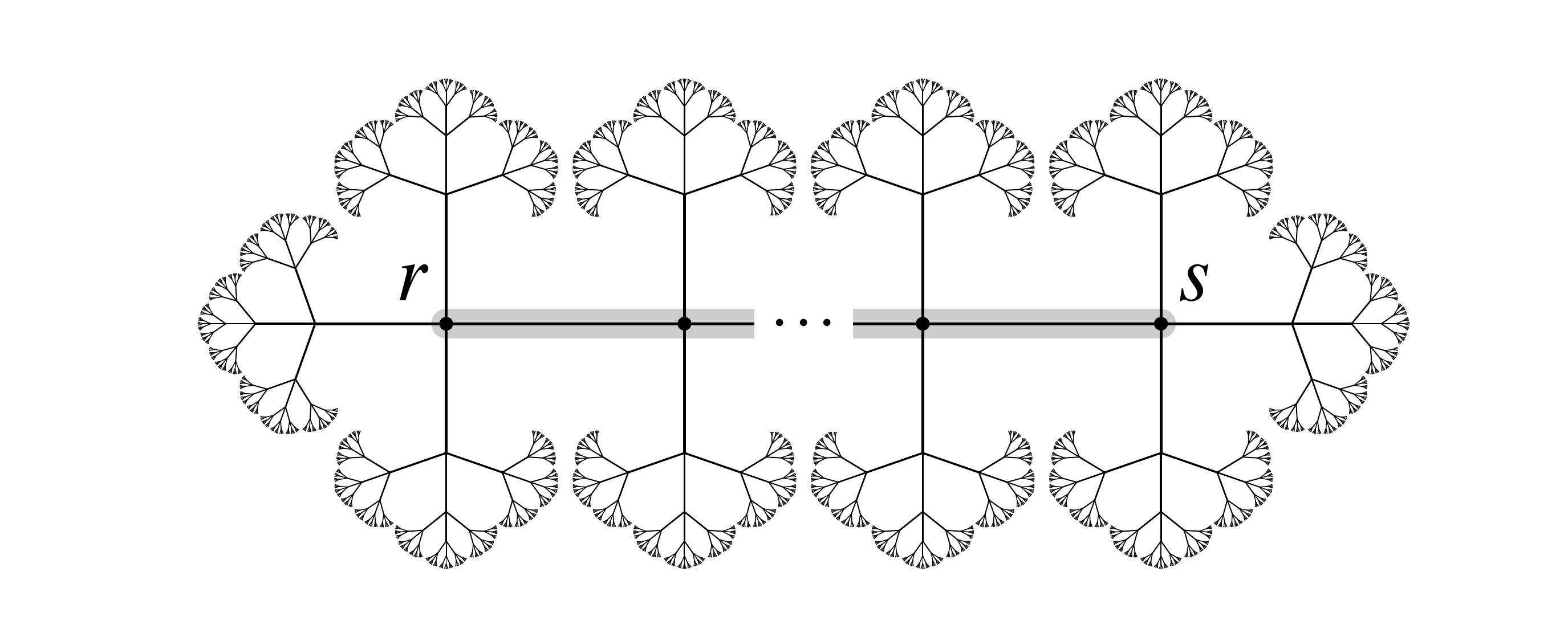}
    \caption{Example of a path from $r$ to $s$ on the Bethe lattice $\mc{B}_4$, which maps to a \emph{nonreversing} walk of length $\ell$ on the square lattice. For any given two sites on the square lattice, there are multiple nonreversing walks that connect the two sites, enumerated by the generating function $C_s(x)$.}
    \label{fig:bethe-rs-path}
\end{figure}

We now generalise this result to include open walks. Consider a walk that begins at site $r$ and ends at site $s$ on $\mc{B}_z$, where $r$ and $s$ are separated by a total of $\ell$ bonds on the Bethe lattice. We denote the corresponding generating function as $T_{rs}(x)$ [by symmetry, $T_{rs}(x) = T_{sr}(x)$]. We will for convenience draw the Bethe lattice as in Fig.~\ref{fig:bethe-rs-path}, the links between $r$ and $s$ (inclusive) forming a
backbone, and refer to $s$ as being to the right of $r$, such that $s = r + \ell$. The walk from $r \to s$ may then be decomposed in the following way:
\begin{enumerate}[(i)]
    \item \begin{enumerate}[(a)]
        \item hopping to one of the $z-1$ nearest neighbours of $r$ not equal to $r+1$,
        \item performing a self-retracing walk that begins and ends at depth $k=1$,
        \item hopping back to $r$,
        \item performing a walk from $r$ to $s$.
    \end{enumerate}
    \item \begin{enumerate}[(a)]
        \item hopping to the `right' of $r$ to site $r+1$,
        \item performing a walk from $r + 1$ to $s$.
    \end{enumerate}
\end{enumerate}
The walk from $r+1$ to $s$ is then decomposed in a similar fashion. Therefore, in terms of the individual generating functions,
\begin{equation}
    T_{rs}^{(z)}(x) = \underbrace{(z-1)x^2 R^{(z)}(x) T_{rs}^{(z)}(x)}_{\text{(i)}} + \underbrace{x T_{r+1 \, s}^{(z)}(x)}_{\text{(ii)}}
    \, ,
    \label{eqn:generating-function-recursion-relation}
\end{equation}
where $R^{(z)}(x)$ is the generating function for self-retracing walks that begin and end at a depth of $k \geq 1$. The labels (i) and (ii) refer to the corresponding steps in the above physical decomposition. From our previous analysis of $T_0^{(z)}(x)$, we know that
\begin{equation}
    R^{(z)}(x) = \cfrac{1}{1 - \cfrac{(z-1)x^2}{1- \cfrac{(z-1)x^2}{1-\ddots}}} = \frac{1-\sqrt{1-4(z-1)x^2}}{2(z-1)x^2}
    \, ,
\end{equation}
where again the sign of the square root is chosen to give $R^{(z)}(x) \to 1$ in the limit $x \to 0$. The recursion relation \eqref{eqn:generating-function-recursion-relation} can then be solved to find an expression for $T_{rs}^{(z)}(x)$ in closed form:
\begin{equation}
    T_{rs}^{(z)}(x) = \left[ x R^{(z)}(x) \right]^\ell T_{ss}^{(z)}(x)
    \, .
\end{equation}
The function $T_{ss}^{(z)}(x)$ which terminates the recurrence relation is simply $T_0^{(z)}(x)$ derived in the previous section, i.e., enumerating the number of perfectly self-retracing paths that begin and end at the same point on the Bethe lattice. We therefore arrive at the final result:
\begin{align}
    T_{rs}^{(z)}(x) &= \left( \frac{1-\sqrt{1-4(z-1)x^2}}{2(z-1)x} \right)^\ell
    \frac{2(z-1)}{z-2+z\sqrt{1-4(z-1)x^2}}\\
    &\equiv S^{(z)}(x)^\ell \, T^{(z)}(x)
    \, .
    \label{eqn:general-self-retracing-GF}
\end{align}
By virtue of the symmetry of the Bethe lattice, $T_{rs}^{(z)}(x)$ depends only on the length $\ell$ of the path separating the sites $r$ and $s$, not on the specific choice of
path.
%
%

\subsection{Constrained closed walks}

We now further generalise to the case of closed, self-retracing walks on the original lattice $\mc{L}_z$ on which the spinon hops in real space.
In order to calculate $P_s(t)$, we are required to enumerate the number of perfectly self-retracing paths that visit the sites $0 \to s \to 0$.
Any such path can be decomposed as follows:
\begin{enumerate}[(i)]
    \item a \emph{nonreversing} base path connecting $0$ and $s$ on $\mc{L}_z$,
    \item self-retracing excursions which decorate the base path.
\end{enumerate}
The base path must be common to both outward ($0 \to s$) and return ($s \to 0$) paths, while the self-retracing excursions can differ between the two paths.
In this way, the return path completely ``erases'' the outwards path, and the path is overall perfectly self-retracing, therefore enclosing precisely zero area.
The base paths must be nonreversing, since immediate reversal of the base
path corresponds to a self-retracing excursion, which would lead to double
counting of such a path.
An example of a self-retracing round
trip between two sites is shown in Fig.~\ref{fig:bethe-square-mapping}.

The connection between the Bethe lattice and the original lattice comes from the number of base paths that the particle may take to get between the origin and the site $s$.
Suppose that we know the generating function for the number of nonreversing paths that connect the origin ($0$), and some other site $s$ \emph{on the original lattice} $\mc{L}_z$, which we denote by
\begin{equation}
    C_s(x) = \sum_{\ell=0}^{\infty} c^{(s)}_\ell x^\ell
    \, .
    \label{eqn:2d-lattice-paths-GF}
\end{equation}
The generating function for fully self-retracing paths that connect $0 \to s \to 0$ can then be constructed in the following way from the three elementary generating functions $S^{(z)}(x)$, $T^{(z)}(x)$ and $C_s(x)$, defined in \eqref{eqn:general-self-retracing-GF} and \eqref{eqn:2d-lattice-paths-GF}. We will henceforth drop the
explicit dependence of these generating functions on the coordination
number $z$ for notational convenience.
Using generating variables $x$ and $y$ to count the number of steps taken on the outwards and return trips, respectively,
\begin{align}
    \mathcal{P}_s(x, y) &= \sum_{\ell=0}^\infty c^{(s)}_\ell S(x)^\ell T(x) S(y)^\ell T(y) \\
    &= T(x)T(y) C_s\left[ S(x) S(y) \right] \label{eqn:Ps-generating-function}
    \, ,
\end{align}
i.e., for each base path, and at each step, a self-retracing excursion may take place, enumerated by the functions $S(x)$ and $T(x)$.
As required, the generating function $\mathcal{P}_s(x, y)$ is symmetric under exchange of forwards and backwards walks (i.e., $x \leftrightarrow y$).
%
%

\subsection{Nonreversing walks}

Equation~\eqref{eqn:Ps-generating-function} shows that the number of nonreversing base paths on the original lattice, enumerated by $C_s(x)$, is a crucial ingredient in determining the transition probability $P_s(t)$.
Our final task therefore is to determine explicitly the generating function $C_s(x)$ (for an arbitrary site $s$), a general method for which is presented here.
We will introduce the strategy for the square lattice, with the generalisation to the triangular and honeycomb lattices (relating to the quasiparticle excitations on the kagome lattice) are deferred to Appendix~\ref{sec:other-lattices}.

Since the nonreversing constraint only depends on the previous step in the lattice walk, it may be enforced using $z \times z$ matrices~\cite{Temperley1956}.
Let us introduce the generating variables $x$, $\delta$ and $\epsilon$ which count the length of the walk, and the number of steps taken in the direction of the (for the square lattice, orthonormal) lattice vectors $\v{e}_1$ and $\v{e}_2$, respectively. At each step, there are four possible directions that the particle may choose from: $\delta$, $\epsilon$, $\epsilon^{-1}$ and $\delta^{-1}$.
However, for all but the initial step of the walk, the direction which immediately reverses the previous step is forbidden.
This may be enforced using the matrix
\begin{equation}
    N = x
    \begin{pmatrix}
        \delta & \epsilon & \epsilon^{-1} & 0 \\
        \delta & \epsilon & 0 & \delta^{-1} \\
        \delta & 0 & \epsilon^{-1} & \delta^{-1} \\
        0 & \epsilon & \epsilon^{-1} & \delta^{-1}
    \end{pmatrix}
    \, ,
\end{equation}
and the initial condition $N_0 = x \diag(\delta, \epsilon, \epsilon^{-1}, \delta^{-1})$.
The row index corresponds to the previous step, and the column index to the current step.
At each step, the length of the path is advanced by one, and matrix multiplication ensures that all possible combinations of steps are accounted for.
The zero entries enforce the nonreversing constraint---any path that immediately reverses its direction is given a coefficient of zero.
The initial matrix $N_0$ imposes that the initial step is unconstrained.
Thence the elements of the matrix $N_0 N^{\ell - 1}$ give the paths of length $\ell$ that are consistent with the nonreversing constraint.
The full generating function $\mc{N}(x; \delta, \epsilon)$ for nonreversing paths is therefore given by the sum over all matrix elements and all possible path lengths $\ell$ (including also the trivial walk of zero length):
\begin{align}
    \mc{N}(x; \delta, \epsilon) &= 1 + \sum_{i,j}\sum_{\ell = 1}^\infty \left[ N_0 N^{\ell - 1}\right]_{ij} \\
    &= 1 + \sum_{i,j} \left[ N_0 (\mathds{1}_z-N)^{-1} \right]_{ij}
    \, ,
    \label{eqn:nonreversing-gf-general}
\end{align}
where $\mathds{1}_z$ is the $z \times z$ identity matrix. Evaluating the inverse of the matrix $\mathds{1}_z-N$, we arrive at the following explicit expression for the generating function for nonreversing walks on the square lattice
\begin{equation}
    \mc{N}(x; \delta, \epsilon) = \frac{1-x^2}{1+3x^2 - x (\delta + \delta^{-1} + \epsilon + \epsilon^{-1})}
    \, ,
\end{equation}
consistent with Ref.~\onlinecite{Temperley1956}. This generating function and its counterparts for the other two-dimensional lattices considered in Appendix~\ref{sec:other-lattices} represent a central object in this work since they give access to the family of generating functions $C_s(x)$ for all sites $s$, and hence contain complete information about the spinon density profile after a quench in the magnetic field strength.

Noting that
\begin{equation}
    \mc{N}(x; \delta, \epsilon) = \sum_{s \in \mc{L}_z} \delta^{s_1} \epsilon^{s_2} C_s(x)
    \, ,
    \label{eqn:N-vs-C}
\end{equation}
the function $C_s(x)$, with $s=\sum_i s_i\v{e}_i$, may be extracted from $\mathcal{N}$ by singling out the terms in~\eqref{eqn:N-vs-C} proportional to $\delta^{s_1} \epsilon^{s_2}$. This may be accomplished using the transformation
\begin{equation}
    C_s(x) = \int_{-\pi}^\pi \frac{\mathrm{d}\theta}{2\pi}\, \int_{-\pi}^\pi \frac{\mathrm{d}\phi}{2\pi}\, \mc{N}(x; e^{i\theta}, e^{i\phi}) e^{-is_1\theta -is_2 \phi}
    \, .
\end{equation}
Substituting in for the generating function $\mc{N}(x; \delta , \epsilon)$, we arrive at the following simplified expression
\begin{equation}
    C_s(x) = \frac{(1-x^2)}{2\pi^2 x} \int_0^\pi \mathrm{d}\theta \, \int_0^\pi  \mathrm{d}\phi \, \frac{ \cos(s_1 \theta) \cos(s_2 \phi)}{t - \cos\theta - \cos \phi}
    \, ,
    \label{eqn:non-reversing-GF-integrated}
\end{equation}
where we have defined $t = (1+3x^2)/2x$. This integral may be evaluated by exploiting an equivalence with the Green's function of two dimensional tight-binding models with Hamiltonian $H$. Consider
\begin{equation}
    G(w) = \sum_\v{k} \frac{\ket{\v{k}}\bra{\v{k}}}{w - E(\v{k})}
    \, ,
\end{equation}
which satisfies $(w-H)G=\mathds{1}$. The states $\ket*{\v{k}}$ are eigenstates of $H$ with energies $E(\v{k})$. Taking matrix elements of $G(w)$ with respect to sites $\ket{\v{l}}$, $\ket{\v{m}}$,
\begin{align}
    G(w; \v{l}, \v{m}) &\equiv \mel*{\v{l}}{G(w)}{\v{m}} \\
    &= \frac{1}{\pi^2} \iint_0^\pi \, \frac{ \prod_i \mathrm{d}k_i \cos[k_i (l_i - m_i)]}{w - E(\v{k})}
    \, .
    \label{eqn:2D-tight-binding-greens-function}
\end{align}
Hence, when $H$ corresponds to a two-dimensional tight-binding model on the square lattice with $E(\v{k}) = \cos k_x + \cos k_y$, we observe the equivalence of \eqref{eqn:non-reversing-GF-integrated} and \eqref{eqn:2D-tight-binding-greens-function} up to prefactors, making the identifications $(k_1, k_2) \leftrightarrow  (\theta, \phi)$, $w \leftrightarrow t$ and $s_i \leftrightarrow l_i - m_i$.

As shown in, e.g., Refs.~\onlinecite{Morita1971,Economou2013}, the Green's function $G(w; \v{0}, \v{0})$, which is related to the spinon return probability $P_0(t)$, is given exactly by
\begin{equation}
    G(w; \v{0}, \v{0}) = \frac{2}{\pi w} K\left( \frac{2}{w} \right)
    \, ,
\end{equation}
where $K$ is the complete elliptic integral of the first kind. This result gives rise to the generating function
\begin{equation}
    C_0(x) = \frac{2}{\pi} \left( \frac{1-x^2}{1+3x^2} \right) K \left( \frac{4x}{1+3x^2} \right)
    \, .
    \label{eqn:origin-GF}
\end{equation}
The Green's functions for general sites $\v{l}$, $\v{m}$ (and therefore $C_s$ for a general site $s$) can also be obtained explicitly using the recursion relations presented in Refs.~\onlinecite{Morita1971,Economou2013}.
This procedure is used later in Sec.~\ref{sec:density-short-time} to construct the
spatially-resolved spinon density profile.
%
%

\section{Analytical results}
\label{sec:results}

Now that we have presented all of the preliminary results, we focus on understanding the high-temperature limits of the physical quantities introduced in Sec.~\ref{sec:motivation} that may be inferred from the generating functions for self-retracing walks.
%
%

\subsection{Single spinon density of states}

As noted in Sec.~\ref{sec:motivation:greens-function}, the single-spinon Green's function $G_{ii}(t)$ may at high temperatures be expressed in terms of the exponential generating function $\tilde{T}^{(z)}(x)$ corresponding to closed walks on the Bethe lattice $\mc{B}_z$ which, by construction, enclose zero area. The exponential generating function can be constructed from the ordinary generating function $T^{(z)}(x)$ derived in
Sec.~\ref{sec:closed-walks} using the transformation
\begin{equation}
    \tilde{T}^{(z)}(x) = \oint_C \frac{\mathrm{d}w}{2\pi i} \frac{e^{xw}}{w} T^{(z)} \left( \frac{1}{w} \right)
    \, .
    \label{eqn:ordinary-to-exponential}
\end{equation}
The contour $C$ can be shrunk around the branch cut in $(1/w)T^{(z)}(1/w)$ that lies along the real axis between $-2\sqrt{z-1} < \Re (w) < 2\sqrt{z-1}$~\footnote{The function $T(w^{-1})/w$ does not exhibit a pole at $w=0$.}, which gives rise to the expression
\begin{equation}
    G_{ii}^{(z)}(t) = \int_{-2\sqrt{z-1}}^{2\sqrt{z-1}} \frac{\mathrm{d}u}{2\pi} e^{ihtu} \frac{z\sqrt{4(z-1)-u^2}}{z^2 - u^2}
    \, ,
\end{equation}
or, equivalently, to the Brinkman and Rice~\cite{Brinkman1970} density of states
\begin{equation}
    \rho(\omega) =
    \begin{cases}
        \dfrac{z}{2\pi h} \dfrac{\sqrt{4(z-1)-\omega^2/h^2}}{z^2 - \omega^2/h^2} &\text{ for } |\omega| < 2\sqrt{z-1}h \, , \\
        0 &\text{ otherwise} \, ,
    \end{cases}
    \label{eqn:brinkman-dos}
\end{equation}
for single particle excitations.

In the case of the square lattice, for example, this result predicts that the support of $\rho(\omega)$ is narrowed by 13\% from $|\omega| < 4h$ at zero temperature to $|\omega| < 2\sqrt{3}h$ at ``infinite temperature'', and that $\rho(\omega)$ vanishes like $|2\sqrt{3}h \mp \omega|^{1/2}$ at the positive/negative band edge as opposed to a step-like singularity typical for massive, free quantum particles in two dimensions.
The Bethe lattice mapping does not account for the Lifshitz tails, nor any singular behaviour near $\omega=0$ that has been predicted theoretically~\cite{Gade1993, Altland1999lett, Altland1999} and observed numerically~\cite{Furusaki1999} in similar models. These differences arise from the neglect of loop diagrams as shown in Fig.~\ref{fig:paths}---when a lattice path includes a closed loop, there are two ways in which the loop can be traversed, whereas the Bethe lattice approximation leads to a coefficient of one.
Nevertheless, the mapping does capture many of the salient features of the high-temperature density of states. For example, one may use~\eqref{eqn:brinkman-dos} to understand the behaviour of the density of states with increasing temperature for Majorana fermions in the Kitaev honeycomb model~\cite{Nasu2015}.
%
%

\subsection{Single spinon Green's function}

Analogous to the on-site Green's function, the generating function $T_{ij}(x)$ is related (for $i \neq j$) to the off-diagonal matrix elements of the Green's function $G_{ij} = \langle\langle b^{\phantom{\dagger}}_i(t) b^\dagger_j(0) \rangle\rangle$ at high temperature.
As noted previously, such a quantity is not on its own gauge invariant and must be multiplied by the phases corresponding to a given lattice path $\gamma$ connecting the sites $i$ and $j$, i.e.,
\begin{equation}
    G_{ij}^{(z)}(t \, | \, \gamma) \equiv \left\langle\!\left\langle e^{i \sum_{\langle \alpha \beta \rangle \in \gamma } \phi_{\alpha\beta}} b^{\phantom{\dagger}}_i(t) b^\dagger_j(0) \right\rangle\!\right\rangle
    \, .
\end{equation}
Converting to the corresponding exponential generating function using~\eqref{eqn:ordinary-to-exponential}, the contour can again be shrunk around the branch cut on the real axis between $-2\sqrt{z-1} < \Re (w) < 2\sqrt{z-1}$ and we arrive at the expression
\begin{multline}
    G_{ij}^{(z)}(t\,  |\, \gamma) = [4(z-1)]^{1-\ell/2} \int_0^\pi \frac{\mathrm{d}\phi}{2\pi} e^{iht 2\sqrt{z-1} \cos\phi} \\
    \Im\left\{ \frac{e^{i\ell \phi } \sin\phi}{(z-2)\cos\phi - i z \sin\phi}  \right\}
    \, ,
\end{multline}
where $\ell$ is the length of the path $\gamma$ from $i$ to $j$~\footnote{Note that the independence of $G_{ij}(t\, | \, \gamma)$ on the precise choice of path $\gamma$, other than its length $\ell$, is a consequence of the self-retracing path approximation.}.

We note that this quantity also equals the projection of the wave function $\ket{\psi(t)}$ of a particle initially localised at the origin of the Bethe lattice $\mc{B}_z$ onto a site at depth $\ell$, i.e., $\psi_\ell(t) = \bra{\ell}\ket{\psi(t)}$, at zero temperature (where the time evolution is generated by a nearest neighbour tight-binding Hamiltonian).
%
%

\subsection{Spinon density profile}

We now turn to our main result: characterising the density profile of a spinon initially localised at the origin of the original lattice.
The transition probability $P_s(t)$ can be constructed from the generating function $\mc{P}_s(x, y)$ in~\eqref{eqn:Ps-generating-function} via conversion to the corresponding exponential generating function using complex contour integration:
\begin{equation}
    P_s(t) = \oiint \frac{\mathrm{d}w_1}{2\pi i} \frac{\mathrm{d}w_2}{2\pi i} \frac{e^{iht(w_1-w_2)}}{w_1w_2} \mc{P}_s\left(\frac{1}{w_1}, \frac{1}{w_2}\right)
    \, ,
    \label{eqn:general-probability-distribution}
\end{equation}
over sufficiently large circles in both the $w_1$ and $w_2$ complex planes, for example.
The moments of the probability distribution $P_s(t)$ can then be constructed using the generating functions $T(x)$, $S(x)$ and $C_s(x)$ from Sec.~\ref{sec:generating-functions}:
\begin{align}
    \langle \v{r}^{2k}(t) \rangle &= \oiint \frac{\mathrm{d}w_1}{2\pi i} \frac{\mathrm{d}w_2}{2\pi i} \frac{e^{iht(w_1-w_2)}}{w_1w_2} T\left( \frac{1}{w_1} \right) T\left( \frac{1}{w_2} \right)\nonumber \\
    &\times  \mc{R}_{2k}\left[ S\left( \frac{1}{w_1} \right)S\left( \frac{1}{w_2} \right) \right]
    \, ,
    \label{eqn:RMS-displacement-z}
\end{align}
where we have defined the function
\begin{equation}
    \mc{R}_{2k}(x) = \sum_{s \in \mc{L}_z} \dist(0, s)^{2k} C_s(x)
    \, ,
\end{equation}
with $\dist(0, s) = \sqrt{s_x^2 + s_y^2}$ on the square lattice.
Using the relationship~\eqref{eqn:N-vs-C} between the generating function for nonreversing walks, $\mc{N}(x; \delta, \epsilon)$, and $C_s(x)$, we deduce that the function $\mc{R}_{2k}$ may be expressed in terms of appropriate derivatives of $\mc{N}$:
\begin{align}
    \mc{R}_{2k}(x) &\equiv \sum_{s \in \mc{L}_z} \left(s_x^2 + s_y^2\right)^k C_s(x) \label{eqn:C-weighted-average} \\
    &=  \left\{ \left[ \left( \delta \partial_\delta \right)^2 + \left( \epsilon \partial_\epsilon \right)^2  \right]^k \mc{N}(x; \delta, \epsilon) \right\} \bigg\rvert_{\delta=\epsilon=1}
    \label{eqn:R_2k-vs-N}
    \, ,
\end{align}
which we will write symbolically as $\mc{R}_{2k} = [\grad^{2k}\mc{N}](x; 1, 1)$.
The expression~\eqref{eqn:C-weighted-average} and hence~\eqref{eqn:R_2k-vs-N} must be generalised to include cross-terms between $\delta$ and $\epsilon$ if the two basis vectors $\v{e}_i$ are not orthonormal, as is the case for the triangular and honeycomb lattices (see Appendix~\ref{sec:other-lattices} for further details).
%
%

\subsubsection{Short-time dynamics}
\label{sec:density-short-time}

\begin{figure}[t]
    \centering
    \begin{minipage}{0.065\linewidth}
      \includegraphics[width=\linewidth]{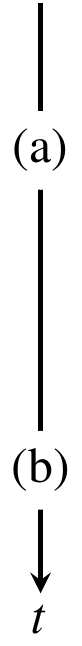}
    \end{minipage}%
    \begin{minipage}{0.935\linewidth}
    \includegraphics[width=0.28\linewidth, valign=center]{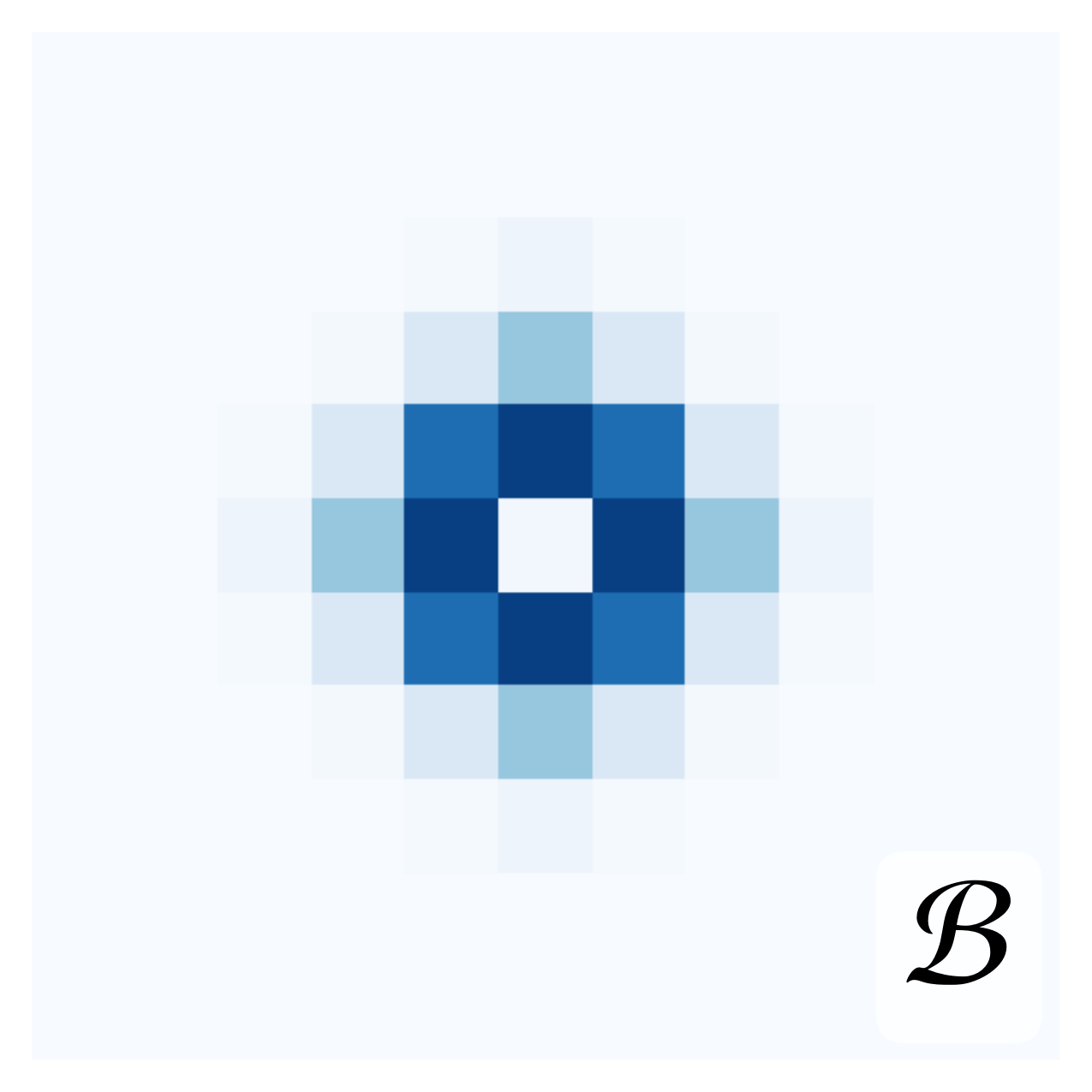}
    \includegraphics[width=0.28\linewidth, valign=center]{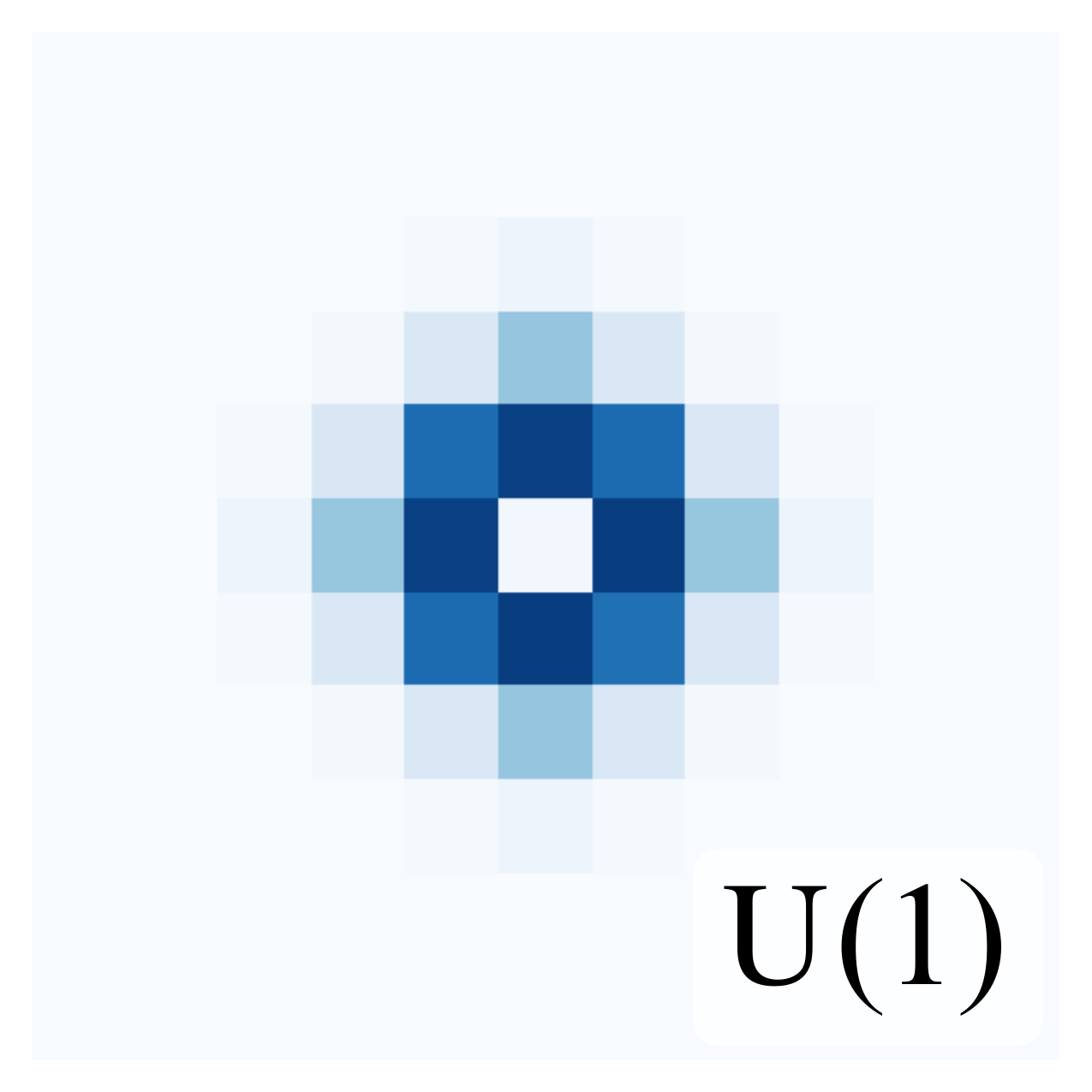}
    \raisebox{-3pt}{\includegraphics[width=0.4\linewidth, valign=center]{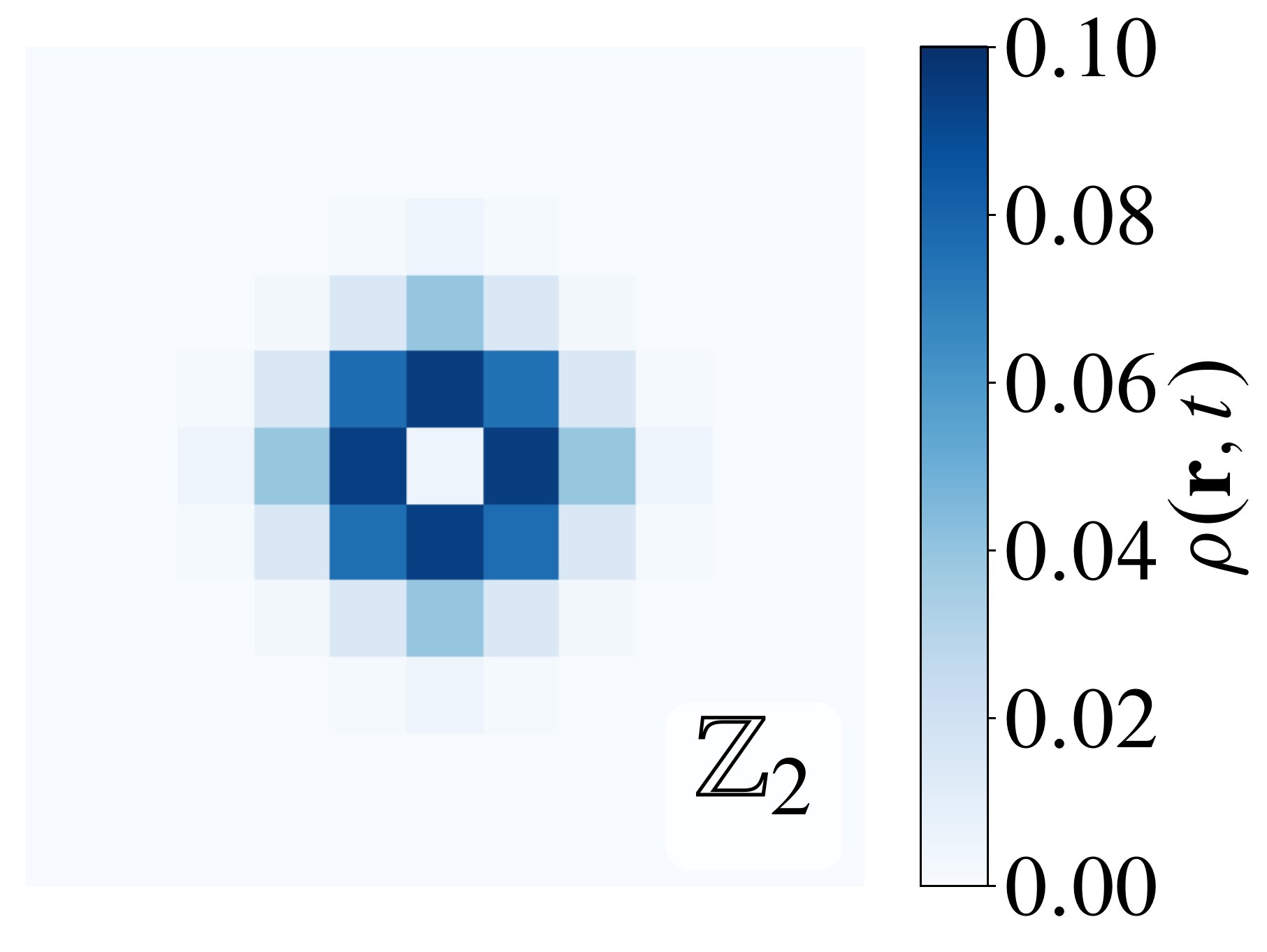}}\\
    \includegraphics[width=0.28\linewidth, valign=center]{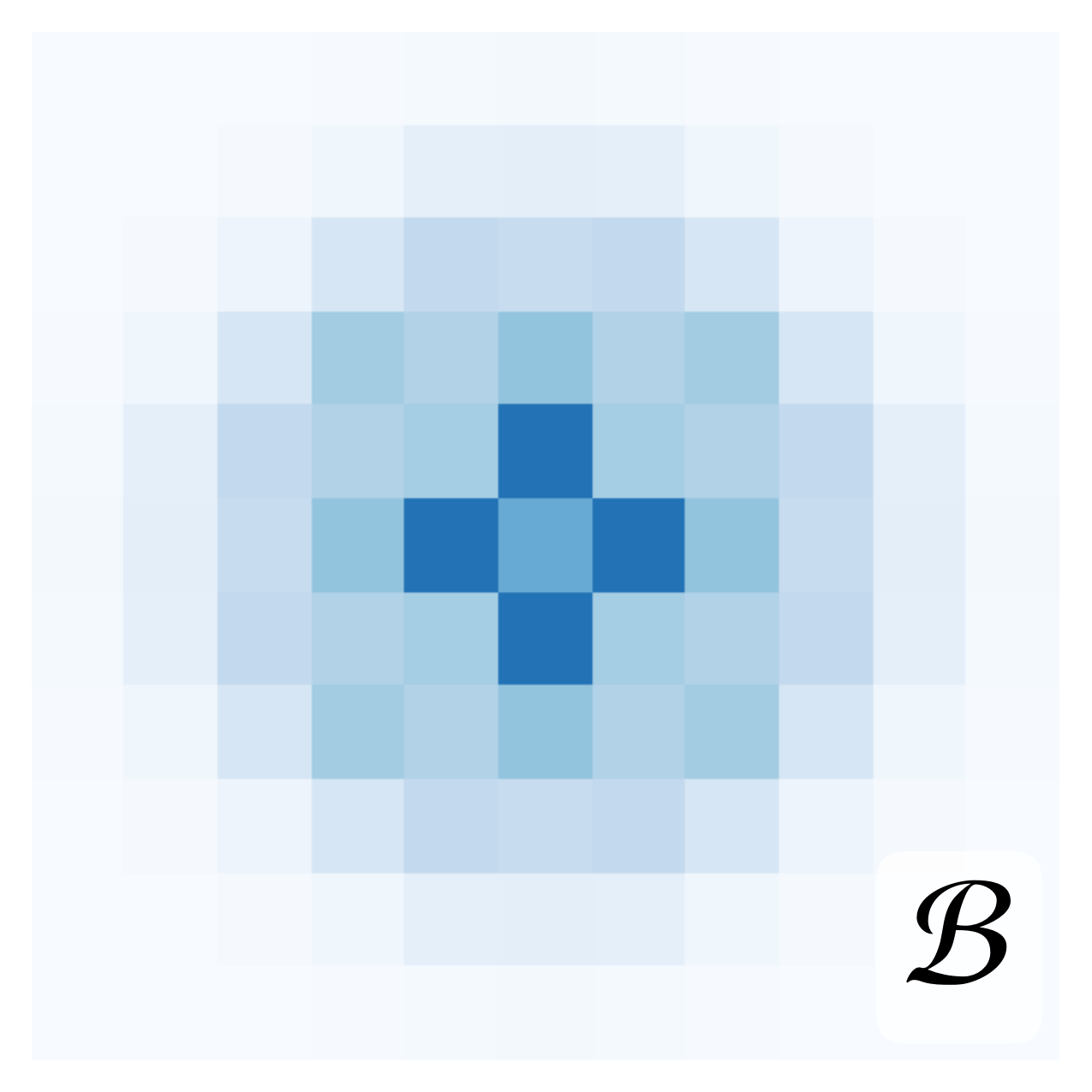}
    \includegraphics[width=0.28\linewidth, valign=center]{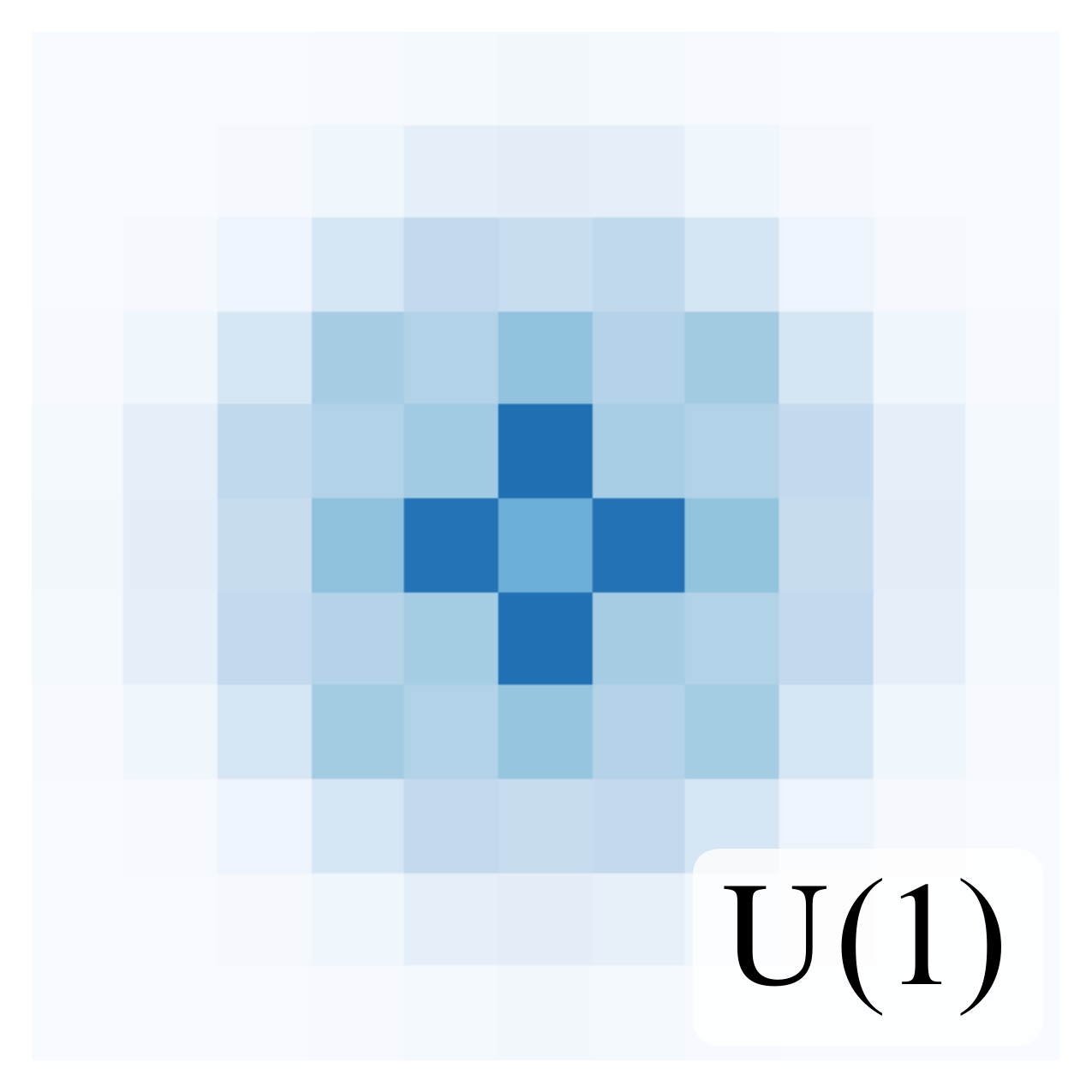}
    \raisebox{-3pt}{\includegraphics[width=0.4\linewidth, valign=center]{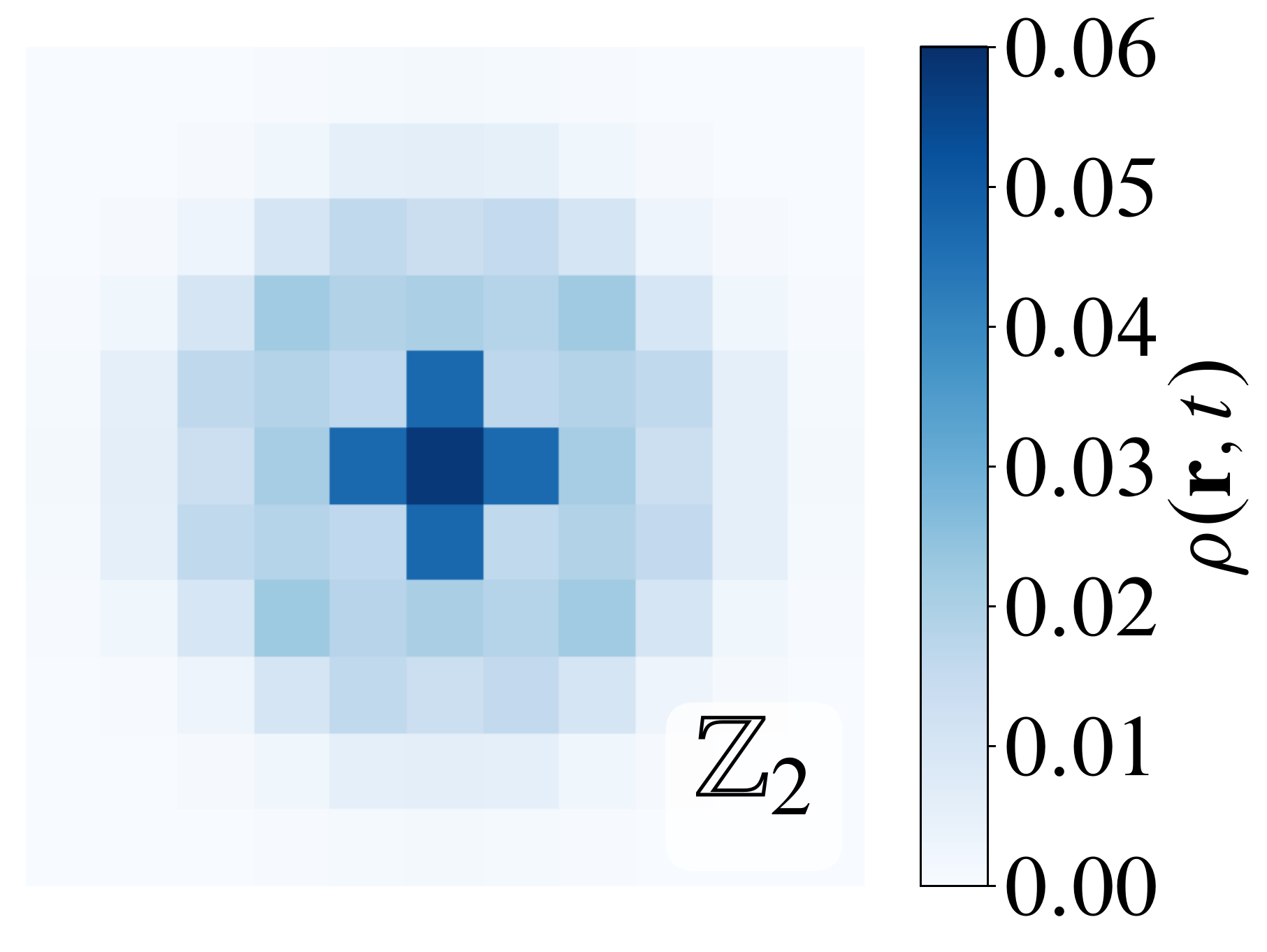}}
  \end{minipage}
    \caption{At short times, the transition probabilities $P_s(t)$ exhibit coherent oscillatory behaviour. We plot the spinon density profile at times (a) $ht=0.9$ and (b) $ht=1.8$, which correspond approximately to a minimum and a maximum of the return probability $P_0(t)$, as predicted by~\eqref{eqn:general-probability-distribution}, respectively. From left to right, the profiles correspond to the Bethe lattice analytical result, the continuous flux model, and the $\pi$-flux model, labelled $\mathcal{B}$, $\text{U}(1)$, and $\mathbb{Z}_2$, respectively. The Bethe lattice mapping provides an essentially exact description of the full density profile for the continuous flux model, and a very good approximation to the $\pi$-flux model, at times on the order of the hopping timescale.
    Notably, discrepancies first become manifest in the $\pi$-flux model at the origin due to the stronger localisation.
    The numerical data for the \Z2 and \U1 cases are averaged over $\num{25000}$ histories.}
    \label{fig:spinon-profile}
\end{figure}
At sufficiently short times, $ht \lesssim \dist(0, s)$, the transition probabilities $P_s(t)$ exhibit complex oscillatory behaviour arising from the interference of lattice walks of varying lengths, and one must use the full expression~\eqref{eqn:general-probability-distribution} in order to accurately capture the density profile in this regime.

Take for example the return probability $P_0(t)$ obtained using the generating function $C_0(x)$ in~\eqref{eqn:origin-GF} (discussed in further detail in Appendix~\ref{sec:return-probability}). The probability decays with time asymptotically as $P_0(t) \!\sim\! 1/t$, suggesting that the spinon asymptotically exhibits diffusive behaviour, which we study in more detail in the next section. However, there also exist superimposed, subleading oscillations due to interference effects that decay as $\cos(4\sqrt{3}t)/t^2$, which may be revealed by applying the method of stationary phase to~\eqref{eqn:general-probability-distribution}.

In Fig.~\ref{fig:spinon-profile} we plot the spinon density profile predicted by~\eqref{eqn:general-probability-distribution} at $ht=0.9$ and $ht=1.8$ [corresponding approximately to extrema of the return probability $P_0(t)$, i.e., $ht \simeq n\pi/(4\sqrt{3})$, for integer $n$], making use of the family of generating functions $\{ C_s(x) \}$, and we compare it with numerical simulations of the disordered tight-binding model~\eqref{eqn:two-spinon-hamiltonian-phases} for the case of (i) $\pi$-fluxes, and (ii) continuous fluxes, $\phi \in [0, 2\pi)$. We observe almost perfect agreement between the analytical results and the numerics at the shortest of the two times, whereas the quantitative agreement survives at the later time for the continuous flux model only.
%
%

\subsubsection{Asymptotic second moment}

We now focus on the asymptotic behaviour of the density profile, once the transient, oscillatory behaviour of the distribution has subsided.
Let us restrict our attention briefly to the second moment of the density distribution, $\langle \v{r}^2(t) \rangle$.
For the case of the square lattice, evaluating the derivatives in \eqref{eqn:R_2k-vs-N}, one arrives at
\begin{equation}
    \mc{R}_2(x) = \frac{4x(1+x)}{(1-3x)^2(1-x)}
    \, .
    \label{eqn:R1-generating-functions}
\end{equation}
Crucially, the function $\mc{R}_2(x)$ has a second order pole at $x=(z-1)^{-1}$.
This feature is shared by the other lattices considered in Appendix~\ref{sec:other-lattices}, and dominates the long-time behaviour of the root mean square (RMS) displacement.
In particular, a second order pole in the function $\mc{R}_2$ gives rise to a \emph{linear}, i.e., diffusive, $t$-dependence of $\langle \v{r}^2(t) \rangle \simeq 2D_zht$, for sufficiently large times.
The full time-dependence of $\langle \v{r}^2(t) \rangle$ described by~\eqref{eqn:RMS-displacement-z} corresponds to a crossover from ballistic to diffusive behaviour at a time $ht \sim 1$ (the characteristic time taken for the spinon to hop one lattice spacing). This is because for sufficiently short times the particle has not moved far enough to enclose any flux, and so interference effects do not play a significant role.
The linear time dependence at long times is a direct consequence of the result
\begin{equation}
    \underset{w=0}{\Res} f(w) \frac{e^{itw}}{w^2} = itf(0) + f'(0) \stackrel{t \gg 1}{\sim} itf(0)
    \, ,
    \label{eqn:linear-t-residue}
\end{equation}
if the function $f(w)$ is analytic at $w=0$.
Note that in fact there exists a line of poles along the real axis in~\eqref{eqn:RMS-displacement-z} since $S(u+i0^+)S(u-i0^+)=(z-1)$ for $u \in \mathbb{R}$ and $|u|<2\sqrt{z-1}$.
Expanding the integrand for general $z$ about this singular line, we must integrate over the relevant residues between $-2\sqrt{z-1} < u < 2\sqrt{z-1}$ (where the integrand is singular), which defines the function
\begin{align}
    \mc{F}(z) &\equiv (z-1)\int_{-2\sqrt{z-1}}^{2\sqrt{z-1}} \mathrm{d}u \, \frac{4(z-1)-u^2}{z^2-u^2} \label{eqn:residue-integral} \\
    &= (z-1) \left[ 4\sqrt{z-1} - z\left(\frac{z-2}{z}\right)^2\ln\left(\frac{z+2\sqrt{z-1}}{z-2\sqrt{z-1}}\right) \right]
    \, .
\end{align}
Comparing the large-$t$ asymptotic expansion of Eq.~\eqref{eqn:RMS-displacement-z} with the expected late time behaviour of $\langle \mathbf{r}^2(t)\rangle \sim 2D_z ht$, one therefore obtains the following exact expression for the diffusion constant on a lattice with coordination number $z$
\begin{equation}
    2D_z = \frac{1}{2\pi} \mc{F}(z) \underset{w\to(z-1)^{-1}}{\lim} \left(1-\frac{w}{z-1}\right)^2\mc{R}^{(z)}_2(w)
    \, .
    \label{eqn:general-D}
\end{equation}
Using the expression for $\mc{R}_2$ in~\eqref{eqn:R1-generating-functions}, we finally deduce that
\begin{align}
    D_4 &= \frac{4}{\pi}\left[2\sqrt{3} - \ln(2+\sqrt{3}) \right]
    \label{eqn:exact-diffusion-constant} \\
    &\simeq 2.73383
    \, ,
\end{align}
for the square lattice.
This expression gives the exact value of the diffusion constant observed in, e.g., Ref.~\onlinecite{Kanasz-Nagy2017}, which was previously obtained only numerically.
The values of the diffusion constant for the triangular and honeycomb lattices are given in Tab.~\ref{tab:diffusion-constants}.

\setlength{\tabcolsep}{16.5pt}
\begin{table}[b]
    \begin{tabular}{@{}lcc@{}}
        \toprule
        Lattice    & Coordination number, $z$ & $D_z$   \\ \midrule
        Triangular & 6   & 2.72968 \\
        Square     & 4   & 2.73383 \\
        Honeycomb  & 3   & 3.20977 \\ \bottomrule
    \end{tabular}
    \caption{Values of the diffusion constants $D_z$ obtained by the Bethe lattice mapping corresponding to the long-time behaviour ($ht \gg 1$) of spinons propagating in the high temperature limit ($T > h^4/J^3$).
    }
    \label{tab:diffusion-constants}
\end{table}

\begin{figure}[t]
    \centering
    \includegraphics[width=\linewidth]{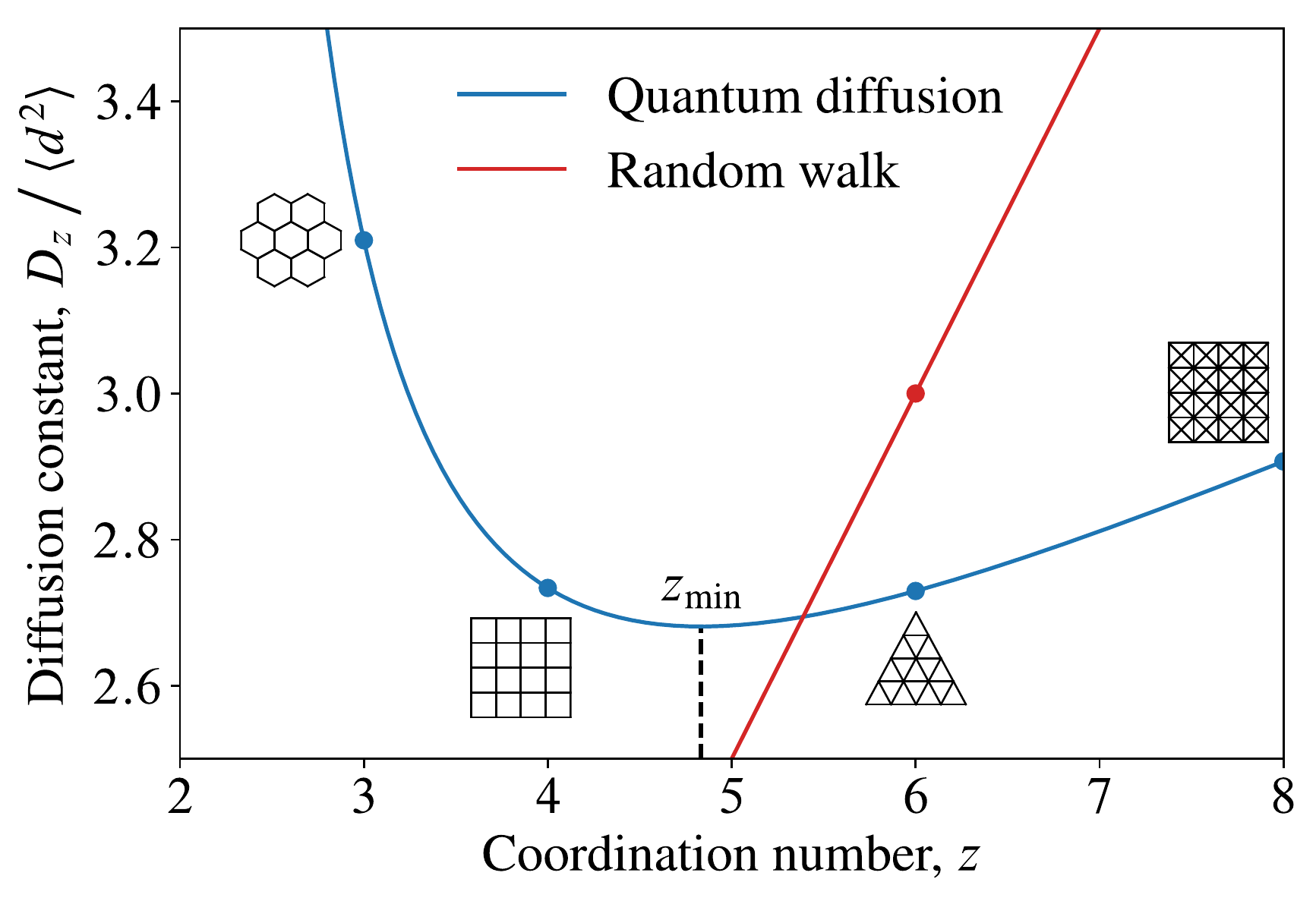}
    \caption{A plot of the diffusion constant $D_z$ against the coordination number $z$ from~\eqref{eqn:general-D_z}, having set the nearest neighbour distance between lattice sites equal to unity.
    The corresponding classical diffusion constant $D_z^{\text{cl}}=z/2$ is also shown for comparison~\cite{classical-diffusion}. The markers denote the values of $D_z$ for some common lattices: the honeycomb, square, triangular, and double triangular lattices, from left to right.}
    \label{fig:D_z}
\end{figure}

In fact, the following expression for $D_z$ is valid for all three lattices
\begin{equation}
    \frac{D_z}{\langle d^2 \rangle} = \frac{z^2}{4\pi (z-2)}\left[ 4\sqrt{z-1} - z\left(\frac{z-2}{z}\right)^2
    \ln\left(\frac{z+2\sqrt{z-1}}{z-2\sqrt{z-1}}\right) \right]
    \, ,
    \label{eqn:general-D_z}
\end{equation}
normalised by the arithmetic mean of the squared distances, $\langle d^2 \rangle$, corresponding to the possible moves at each step~\footnote{This normalisation allows us to apply Eq.~\eqref{eqn:general-D_z} to lattices such as the double triangular lattice ($z=8$), which include steps of different lengths.}.
A plot of this function in Fig.~\ref{fig:D_z} shows that $D_z$ exhibits a minimum at $z \simeq 4.833$---this is due to the competition between (i) reduced destructive interference at low coordination numbers (vanishing destructive interference as $z\to 2^{+}$, since there are no loops for $z=2$), and (ii) a greater number of paths between any two points on the lattice for large $z$, the effect of which dominates at large $z$.
This result shows that it is a fortuitous coincidence that the square and triangular lattices exhibit such similar diffusion constants.

We stress that~\eqref{eqn:general-D_z} corresponds to \emph{quantum} diffusion, which results from the complex interference pattern produced by the multitude of lattice walks, and is faster than the corresponding classical diffusion (random walk), $D_{\text{cl}}= z/2$, for coordination number $z \leq 5$~\cite{classical-diffusion}.
For $z>5$, interference effects dominate, leading to slower propagation.
The difference between classical and quantum diffusion is further reflected in the non-Gaussian nature of the asymptotic density profile, discussed in the next Section.
%
%

\subsubsection{Asymptotic higher order moments}

It is possible to evaluate arbitrary moments of the density distribution in order to give a better characterisation of the spinon density profile.
We specialise to the case of the square lattice (i.e., $z=4$) for convenience.
The function $\mc{R}_{2k}(x)$ in general has a pole of order $k+1$ at $x=(z-1)^{-1}$.
This implies that, in the long-time limit, the $2k$th moment behaves as $\sim\! t^k$, consistent with the diffusive behaviour exhibited by the second moment.
This is because, analogous to~\eqref{eqn:linear-t-residue}, at long times
\begin{equation}
    \underset{w=0}{\operatorname{Res}} \, \frac{f(w)e^{itw}}{w^{k+1}} = \frac{1}{k!}(it)^k f(0) + \ldots
    \, ,
\end{equation}
if $f(w)$ is analytic at $w=0$. The dots correspond to lower powers of $t$, which contribute to the transient oscillatory behaviour at short times.

We start by considering the expression for $\mc{R}_{2k}(x)$ derived in~\eqref{eqn:R_2k-vs-N}
\begin{align}
    \mc{R}_{2k}(x) &= [\grad^{2k}\mc{N}](x; 1, 1) \\
    &= \left[\sum_{\ell=0}^k \binom{k}{\ell} (\delta\partial_\delta)^{2\ell}(\epsilon\partial_\epsilon)^{2(k-\ell)} \mc{N} \right](x; 1, 1)
    \, .
\end{align}
One can show that the term $(\delta\partial_\delta)^{2\ell}(\epsilon\partial_\epsilon)^{2(k-\ell)} \mc{N}$ gives rise to a contribution
\begin{equation}
    \binom{k}{\ell} [2\ell]![2(k-\ell)]! \frac{x^{k}(1+x)}{(1-3x)^{k+1}(1-x)^{k}} + \ldots
\end{equation}
to the highest order pole $\sim\! (1-3x)^{-k-1}$. The dots correspond to poles of lower order that give rise to lower powers of time.
Performing the summation over $\ell$, we obtain
\begin{equation}
    \sum_{\ell=0}^k \binom{k}{\ell}^2 [2\ell]![2(k-\ell)]! = 4^k (k!)^2
    \, .
\end{equation}
Hence, sufficiently close to the pole at $x=1/3$, the function $\mc{R}_{2k}$ behaves as
\begin{equation}
    \mc{R}_{2k}(x) \sim \frac{2^{2+k}}{3} \frac{\Gamma(k+1)^2}{(1-3x)^{k+1}}
    \, ,
\end{equation}
where $\Gamma(x)$ is the Gamma function. The final ingredient therefore is the integral over residues, which generalises the expression in~\eqref{eqn:residue-integral}
\begin{multline}
    \int_{-2\sqrt{3}}^{2\sqrt{3}} \mathrm{d}u \frac{(12-u^2)^{\frac{k+1}{2}}}{16-u^2}
    = \\ \sqrt{\pi} 2^{k-2} 3^{(k+2)/2} \frac{\Gamma\left(\tfrac{3+k}{2}\right)}{\Gamma\left(\tfrac{4+k}{2}\right)} {}_2{F}_1\left( \begin{matrix}\tfrac{1}{2},\,\,\,\, 1 \\ \, 2+\tfrac12 k\, \end{matrix} ; \frac{3}{4} \right)
    \, ,
\end{multline}
in terms of the Gauss hypergeometric function ${}_2{F}_1(a, b, c; z)$. Combining the multiplicity of the highest order pole and the integral over residues, we arrive at the final exact expression for the $2k$th moment of the density distribution in the long-time limit, for fixed $k$:
\begin{align}
    \mu_{2k} &\equiv  \lim_{t\to\infty} \frac{\ev*{\v{r}^{2k}(t)}}{(ht)^k}\\
    &= \frac{2^{2k-1} 3^{k/2 + 1}}{\sqrt{\pi}} \frac{\Gamma(k+1)\Gamma(\frac{3+k}{2})}{\Gamma(\frac{4+k}{2})} {}_2F_1\left( \begin{matrix}\tfrac{1}{2},\,\,\,\, 1 \\ \, 2+\tfrac12 k\, \end{matrix} ; \frac{3}{4} \right)
    \, .
    \label{eqn:general-moments}
\end{align}
These moments are checked against numerics in Appendix~\ref{sec:numerical-verification}.
As required, the special case $k=1$ simply reduces to $2D_4$ given in~\eqref{eqn:exact-diffusion-constant}.
The density distribution is, however, \emph{not} Gaussian, as evidenced by nonzero higher order cumulants.
This is not an artefact of the Bethe lattice mapping, and indeed is reflected in our numerical simulations, as seen in Fig.~\ref{fig:rms-comparison}. This feature further distinguishes the interference-driven quantum diffusion from its classical counterpart.

One can similarly compute the moments of the marginal distribution $\rho(r_x; t) = \int \mathrm{d}r_y \rho(\v{r}; t)$. One must in this case replace $\mc{R}_{2k}(x)$ by the function
\begin{equation}
    \tilde{\mc{R}}_{2k}(x) = [(\delta \partial_\delta)^{2k} \mc{N}](x; 1, 1)
    \, .
\end{equation}
A similar calculation of the residues and multiplicities (now simplified by the removal of cross terms between the generating variables $\delta$ and $\epsilon$) gives
\begin{align}
    \mu_{2k}^x &\equiv \lim_{t\to\infty} \frac{\ev*{r_x^{2k}(t)}}{(ht)^k} \\
    &= \frac{3^{k/2 + 1}}{2\sqrt{\pi}} \frac{\Gamma(2k+1)\Gamma(\frac{3+k}{2})}{\Gamma(k+1)\Gamma(\frac{4+k}{2})} {}_2F_1\left( \begin{matrix}\tfrac{1}{2},\,\,\,\, 1 \\ \, 2+\tfrac12 k\, \end{matrix} ; \frac{3}{4} \right) \label{eqn:marginal-moments-exact} \, .
\end{align}
As one would expect, $\mu_{2}^x = D_z$. The first two of these exact moments were used to construct the analytical estimate of the marginal density profile in Fig.~\ref{fig:rms-comparison}.
%
%

\section{Numerical results}
\label{sec:numerical-results}

\begin{figure}[t]
    \centering
    \begin{minipage}{0.585\linewidth}
        \subfloat{\includegraphics[width=\linewidth]{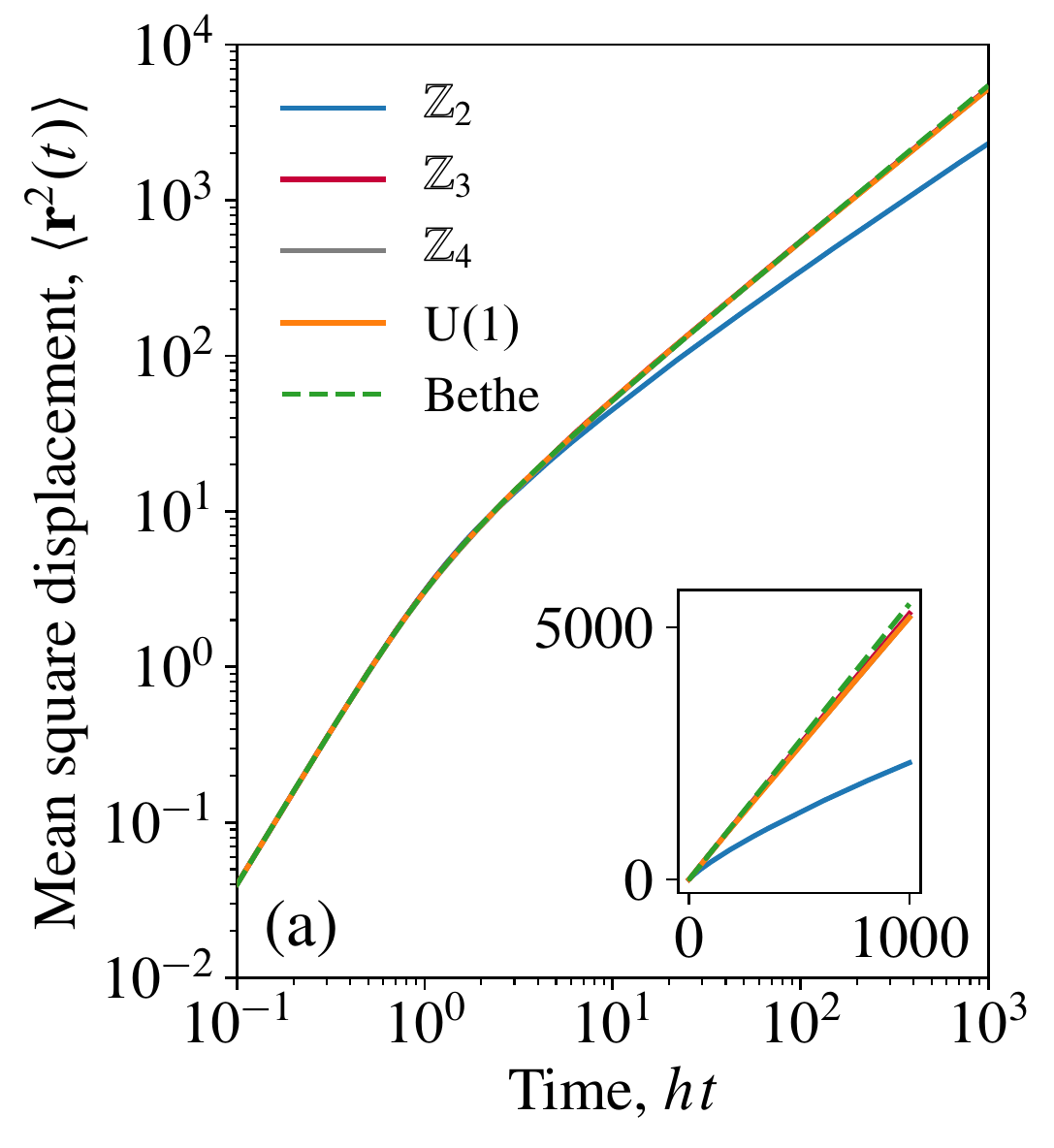}}
    \end{minipage}%
    \begin{minipage}{0.415\linewidth}
        \subfloat{\includegraphics[width=\linewidth]{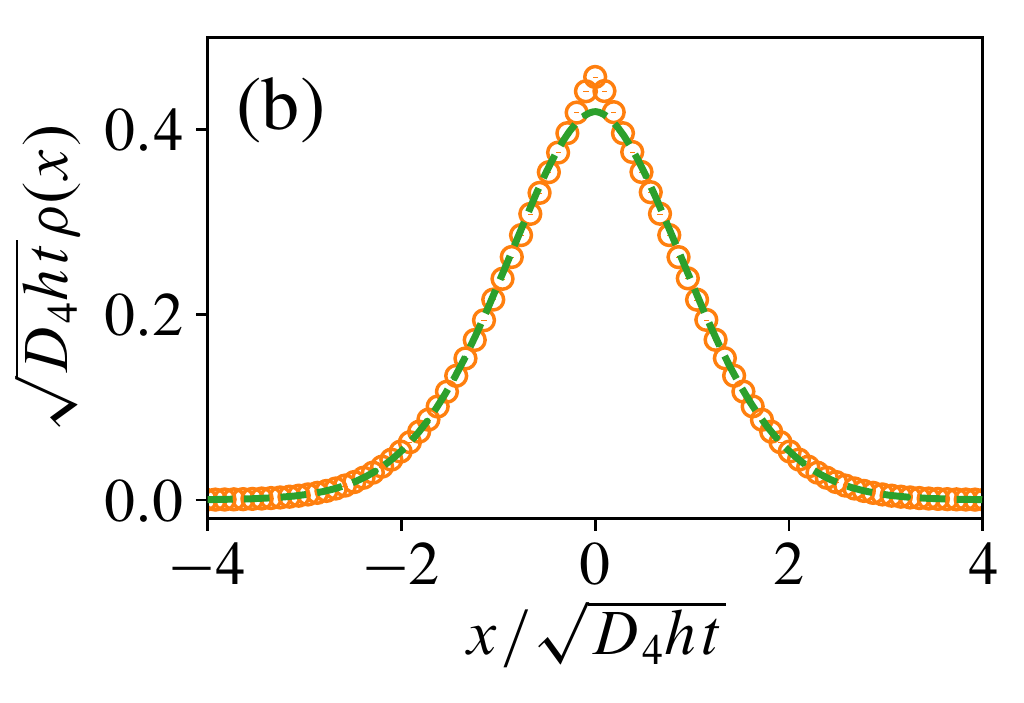}}\\
        \subfloat{\includegraphics[width=\linewidth]{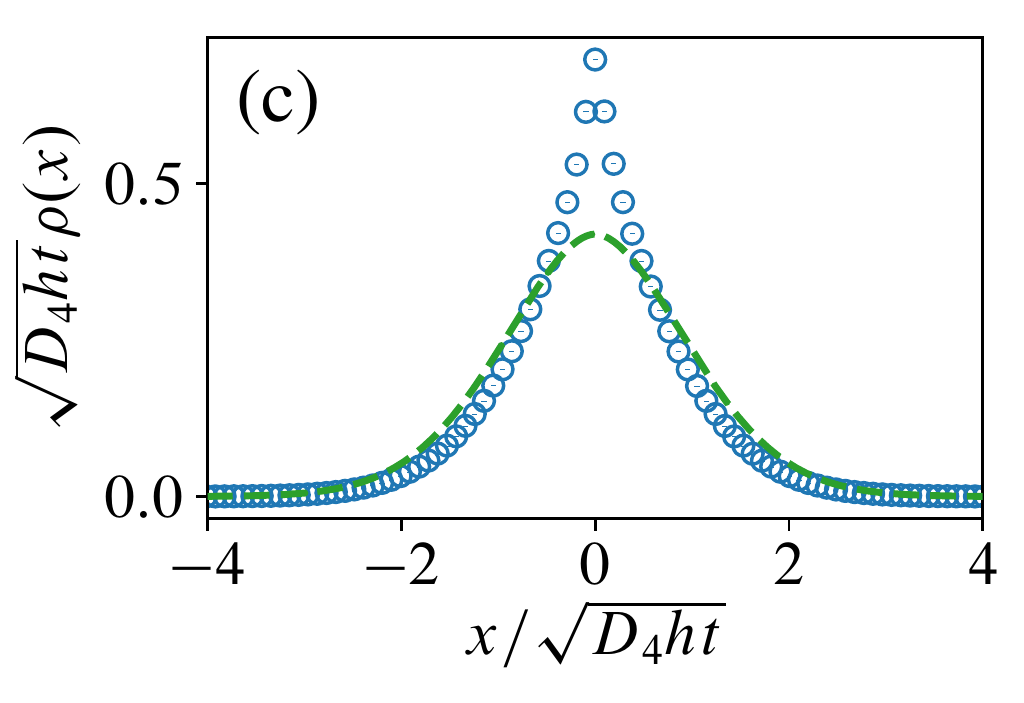}}
    \end{minipage}
    \caption{(a) Comparison of Bethe lattice result for the mean square deviation, $\langle \v{r}^2(t) \rangle \simeq 2D_z ht$, with numerics for (i) the discrete flux model with fluxes $2\pi / n$, for $n=2,3,4$, and (ii) the continuous random flux model.
    Time evolution is performed using a high order Suzuki--Trotter decomposition on a square lattice with $1999 \times 1999$ sites and averaged over $128$ random flux configurations at infinite temperature.
    All models exhibit the same ballistic behaviour for sufficiently short times.
    The $\pi$-flux model shows the most pronounced deviation from the Bethe lattice approximation at longer times as a result of being most strongly localised.
    The density profile at $ht=40$ for the Bethe lattice is compared with the continuous and $\pi$-flux models in (b) and (c), respectively.
    The profiles $\rho(x)$ are averaged over $10^5$ flux configurations on a lattice of size $249 \times 249$ sites.
    The error bars are in all cases too small to be visible.
    }
    \label{fig:rms-comparison}
\end{figure}

The Bethe lattice result is compared to Trotterised time evolution~\cite{DeRaedt1987} on a $\num{1999} \!\times\! \num{1999}$ square lattice, averaged over infinite temperature disorder realisations, both for random discrete fluxes ($2\pi/n$, for $n=2,3,4$), as well as for the continuous random flux model, where the flux threading each plaquette is chosen from a uniform distribution $\phi \in [0, 2\pi)$.
The results are shown in Fig.~\ref{fig:rms-comparison}.
We see that, over numerically accessible, intermediate timescales, the Bethe lattice result provides an excellent quantitative description of the density profile for the continuous flux model,
and qualitative agreement with the \Z2 model.
The results for discrete fluxes with $n>2$ can be seen to rapidly converge to the continuous flux
result.
Note that the case $n=2$, corresponding to \Z2 fluxes, is special, being
the only case in which the effective tight-binding Hamiltonian exhibits
time reversal symmetry (i.e., $H=H^*$).

In the $\pi$-flux model, encircling a flux an even number of times gives rise to constructive interference.
The difference in behaviour between this model and the continuum case means that the loop diagrams depicted in Fig.~\ref{fig:paths:b}, which are missed by the Bethe lattice, play an important role.
These diagrams lead to an increased weight near the origin, which results in a reduced RMS displacement.
These observations are consistent with the idea that all single-particle eigenstates of the effective disordered tight-binding model are localised, but with a diverging localisation length near $E=0$ arising from the presence of purely off-diagonal disorder in the Hamiltonian.

The subdiffusive form of the second moment observed in the numerics may be explained in the following way: a wavepacket composed of states in the vicinity of energy $E$ will diffuse with diffusion constant $D(E)$ up to (approximately) their localisation length $\lambda(E)$~\cite{Kawarabayashi1995}, at which point such states give rise to a fixed, time-independent contribution to $\langle \v{r}^2(t) \rangle \sim \lambda(E)^2$. As time progresses, a reduced fraction of states have not yet reached their localisation length and are still diffusing, explaining the negative curvature observed in Fig.~\ref{fig:rms-comparison}.
The Bethe lattice result can therefore be thought of as giving the behaviour of $\langle \v{r}^2(t) \rangle$ before any of the states have reached their localisation length.
Since the localisation length of the continuous flux model increases exponentially with energy away from the band edge~\cite{Furusaki1999}, we observe very close agreement up to $ht=10^3$ with the Bethe lattice result.
The difference in behaviour between the continuous and $\pi$-flux models may be attributed to the fact that the latter is more strongly localised~\cite{Tadjine2018}, implying a smaller fraction of diffusing states at any given time, and hence a more pronounced departure from the pure diffusion predicted by the Bethe lattice.

\begin{figure}[t]
  \includegraphics[width=\linewidth]{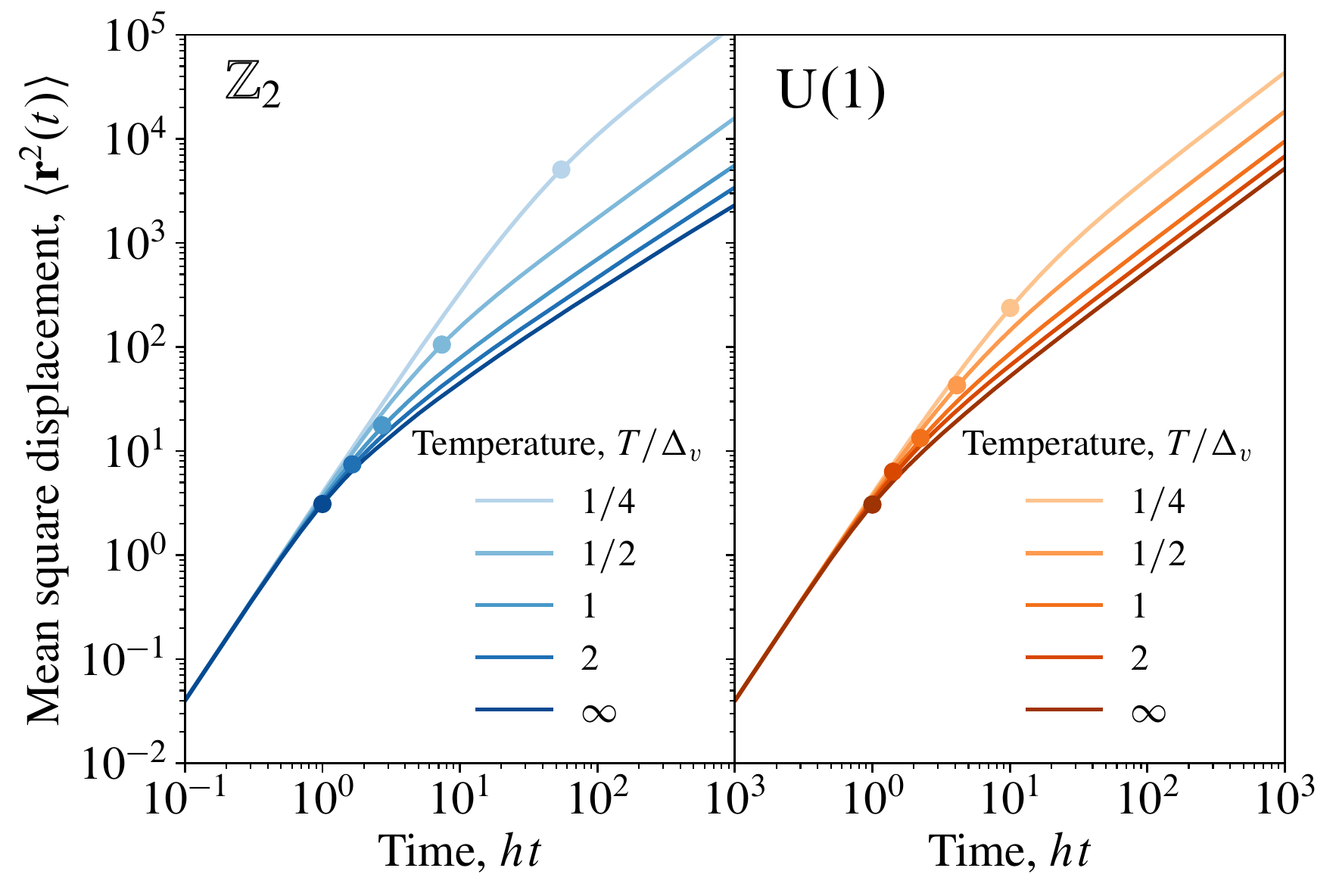}
  \caption{Comparison of the mean square displacement $\langle \v{r}^2(t) \rangle$ of a spinon initially localised at the origin for various temperatures. Over the simulated timescales, the dominant effect of reducing temperature is to shift the crossover from ballistic to (sub-) diffusive behaviour to larger times, namely at a time $ht \sim \xi^2$ indicated by the circular markers as a guide to the eye~\cite{regularisation}.
  In the $\mathbb{Z}_2$ case, $\xi$ is set by the distance between visons. In the $\text{U}(1)$ case, $\xi^2$ is given by the area whose enclosed flux has an $O(1)$ variance. Time evolution is performed using a high order Suzuki--Trotter decomposition on a $1999 \times 1999$ square lattice, and the data are averaged over 128 disorder realisations for each temperature. The error in the data is smaller than the line width.}
  \label{fig:finite-T}
\end{figure}
%
%

%

\subsection{Finite temperature}

In the case of discrete $\pi$-fluxes with a gap $\Delta_v \!\sim\! h^4/J^3$, at intermediate temperatures or, equivalently, finite vison separation $\xi \!\sim\! \rho^{-1/2} \!\sim\! e^{\Delta_v/2T}$, the crossover from ballistic to subdiffusive behavior
is shifted to later times. We expect that the particle
should propagate ballistically until it has encountered
a sufficient number of visons so as to impede its motion:
$(ht)^2/\xi^2 \sim ht$.
Hence, for $ht \ll \xi^2$ we expect to see free-particle behaviour, and for $ht \gg \xi^2$ we expect to observe approximately the infinite temperature (sub-) diffusive behaviour (with a renormalised diffusion constant). The crossover between the two regimes is therefore set by the time taken to diffuse to the nearest vison. This behaviour is indeed seen in Fig.~\ref{fig:finite-T}.

In the continuous flux model, one may attribute an energy cost $E(\phi)=-\Delta_v \cos\phi$ to threading a given plaquette with a flux $\phi$.
At sufficiently low temperatures, $\beta \Delta_v \gg 1$, the corresponding probability density $p(\phi) \propto e^{-\beta E(\phi)}$ is approximately Gaussian, and the relevant length scale $\xi_c$ becomes  $\xi_c^2(T)=2/T$ [cf.~Eq.~\eqref{eqn:vison-separation-lengthscale}]. This characteristic area is defined via the relation $\langle e^{i\sum_{\langle\alpha\beta\rangle \in \gamma} \phi_{\alpha\beta}} \rangle \equiv e^{-A_\gamma / \xi_c^2}$, and may be understood as the area such that typical fluctuations of the enclosed flux have a magnitude that is $O(1)$. As in the discrete flux case, the effect of finite temperature is to shift the crossover from ballistic to (sub-)diffusive behaviour to a time $ht \sim \xi_c^2$, as shown in Fig.~\ref{fig:finite-T}.
%
%

\section{Conclusions}
\label{sec:conclusions}

In this manuscript we studied the effects of nontrivial mutual statistics
on the propagation of quasiparticles in topological systems at finite temperature. Specifically, we considered a temperature regime where one species of quasiparticle is thermally excited and provides a static  (\`{a} la Born--Oppenheimer) stochastic background for the other species, which are sparse and hop coherently across the lattice. This is a regime of experimental interest in topological quantum spin liquids, where a large separation of energy scales between different species of quasiparticle arises naturally in many realistic model Hamiltonians.

We used a combination of numerical and analytical approaches to investigate toric-code-inspired toy models, where the excitations (dubbed spinons and visons) have anyonic mutual statistics. The effect of nonzero temperature in our model is to populate a finite density of static visons. Due to the mutual statistics of the quasiparticles, visons act as Aharonov--Bohm half flux quanta for the spinons. Within perturbation theory, our model permits an effective description in which the spinons evolve in time according to a two-dimensional tight-binding Hamiltonian in the presence of randomly placed fluxes. Changing temperature alters the density of the fluxes, which, in turn, changes the strength of off-diagonal disorder in the tight-binding Hamiltonian. We also considered models in which the flux threading each plaquette is a multiple of $1/3$ or $1/4$ of the flux quantum, and the case in which the flux is distributed continuously.

Various time-dependent observables for lattice systems, including the spinon density profile in our effective tight-binding description, may be computed by counting discrete lattice paths. In order to make analytical progress, we considered the self-retracing path approximation. Such paths are expected to dominate at intermediate times due to interference effects by virtue of the Aharonov--Bohm effect. To this end, we map the self-retracing paths to walks on an auxiliary Bethe lattice and enumerate such walks exactly. This gives us access to analytical expressions for the spinon density profile as a function of space and time.

For sufficiently short times, namely on the order of the hopping timescale, $\tau$, the density exhibits oscillatory behaviour due to coherent interference effects. On these timescales, the self-retracing path approximation is essentially exact and our results are almost indistinguishable from numerical simulations. At times much greater than the hopping timescale, $t \gg \tau$, the self-retracing path approximation predicts asymptotic quantum diffusive behaviour of the spinon, i.e., $\langle \v{r}^2(t) \rangle \simeq 2 D_z t/\tau$. We obtained an exact expression for the corresponding diffusion constant $D_z$, which depends on the coordination number of the underlying lattice.
The function $D_z$ exhibits a minimum at $z \simeq 5$, where the effects of (i) reduced destructive interference at low coordination numbers, and (ii) an increasing number of paths connecting any two sites at larger coordination numbers, balance one another. The higher moments of the density distribution in the large-time limit exhibit non-Gaussian behaviour, which highlights the difference between quantum and classical diffusion.

Comparison with numerical simulations reveals excellent agreement with the continuous flux model up to $O(10^3)$ hopping times, while for the $\pi$-flux model discrepancies become apparent at much shorter times. This difference is understood as arising from the distinct localisation properties of the two models. When considering $\langle \v{r}^2(t) \rangle$, states with a given energy will diffuse with some characteristic diffusion constant until the corresponding localisation length is reached. The $\pi$-flux model is more strongly localised and so at any given time a larger fraction of states have reached their localisation length, and give rise thereafter to a time-independent contribution to $\langle \v{r}^2(t) \rangle$.

The results that we have presented provide us with a quantitative understanding of the crossover from ballistic to quantum (sub-) diffusive motion of spinons through a sea of thermally-excited visons, which is a direct consequence of their nontrivial mutual statistics. More generally, our work represents a step forward in understanding the dynamics of quantum spin liquids at finite temperature, which is essential to interpret both the relevant experiments and numerical data. Our results demonstrate another way in which the mutual semionic statistics of spinons and visons manifests itself in the dynamical properties of spinons; this paves the way for the possible study of such dynamics as an experimentally viable diagnostic tool for anyonic statistics in many-body systems that exhibit topological order.

We expect that our results may be relevant to several interlaced but distinct contexts of many-body physics. On the condensed matter physics front, while realistic Hamiltonians require including further effects, such as possible interactions between quasiparticles and correlations in the spin background, it is nonetheless tempting to point at the recent experimental advances in the study of Kitaev-model-like candidate materials at finite temperature as a possible context where the physics discussed in our work may be relevant and observable~\cite{Takagi2019}. However, to make such connections, some modification of our present framework is necessary in that the spinon dispersion in the Kitaev model~\cite{Kitaev2003} is massless and relativistic, as opposed to the massive and non-relativistic dispersion considered in the present work.

In the context of quantum information and quantum computing, the recent proposal that the toric code and similar $\mathbb{Z}_2$ spin liquid Hamiltonians may be realised using quantum annealers~\cite{Chamon2019}, indeed in the limit explored in our work of a large star constraint and a perturbative transverse field, promises to provide further avenues to benchmark and explore the type of phenomena that we have uncovered, in a convenient and highly tunable setting.

Seen from the ultracold atomic physics perspective, our results also describe quantitatively the motion of holes in real space in the large-$U$, large-$S$ limit of the Hubbard model~\cite{Kanasz-Nagy2017}.
Owing to recent developments in quantum gas microscopy~\cite{Bakr2009, Sherson2010, Haller2015, Cheuk2015, Parsons2015, Edge2015}, our analytical expressions for the site-resolved density profile may thus be probed directly in ultracold atomic experiments. Our calculation extends the self-retracing path approximation used in Ref.~\onlinecite{Carlstrom2016}, showing that it in fact holds to much larger times and distances.

Our work also points at a couple of potentially interesting future directions. Extending the analysis in the present work to string-net models~\cite{Levin2005} may offer access to richer varieties of topological order and anyonic statistics, and an opportunity to classify more generally the resulting dynamics. Perhaps more interesting is the spinons' back action on the visons, which we have so far neglected. In a similar spirit to Ref.~\onlinecite{Kanasz-Nagy2017}, it is plausible that the quantum coherent hopping of spinons may lead to nontrivial correlations in the positions of the visons, which may have other important implications at finite temperature, indicative of quantum spin liquid behaviour~\cite{Hart2019}.

%
%

\begin{acknowledgements}
The authors would like to thank John Chalker, Ben Irwin and Johannes Knolle for useful discussions. This work was supported in part by Engineering and Physical Sciences Research Council (EPSRC) Grants No.~EP/P034616/1 and No.~EP/M007065/1 (CC and OH).
This work was performed using resources provided by the Cambridge Service for Data Driven Discovery (CSD3) operated by the University of Cambridge Research Computing Service (\href{http://www.csd3.cam.ac.uk/}{\color{black}{http://www.csd3.cam.ac.uk/}}), provided by Dell EMC and Intel using Tier-2 funding from the Engineering and Physical Sciences Research Council (capital grant EP/P020259/1), and DiRAC funding from the Science and Technology Facilities Council (\href{www.dirac.ac.uk}{\color{black}{www.dirac.ac.uk}}).
\end{acknowledgements}
%
%
\appendix

\section{Other lattices}
\label{sec:other-lattices}
%
%

\subsection{Toric code on the kagome lattice}

\begin{figure}
    \centering
    \includegraphics[width=0.85\linewidth]{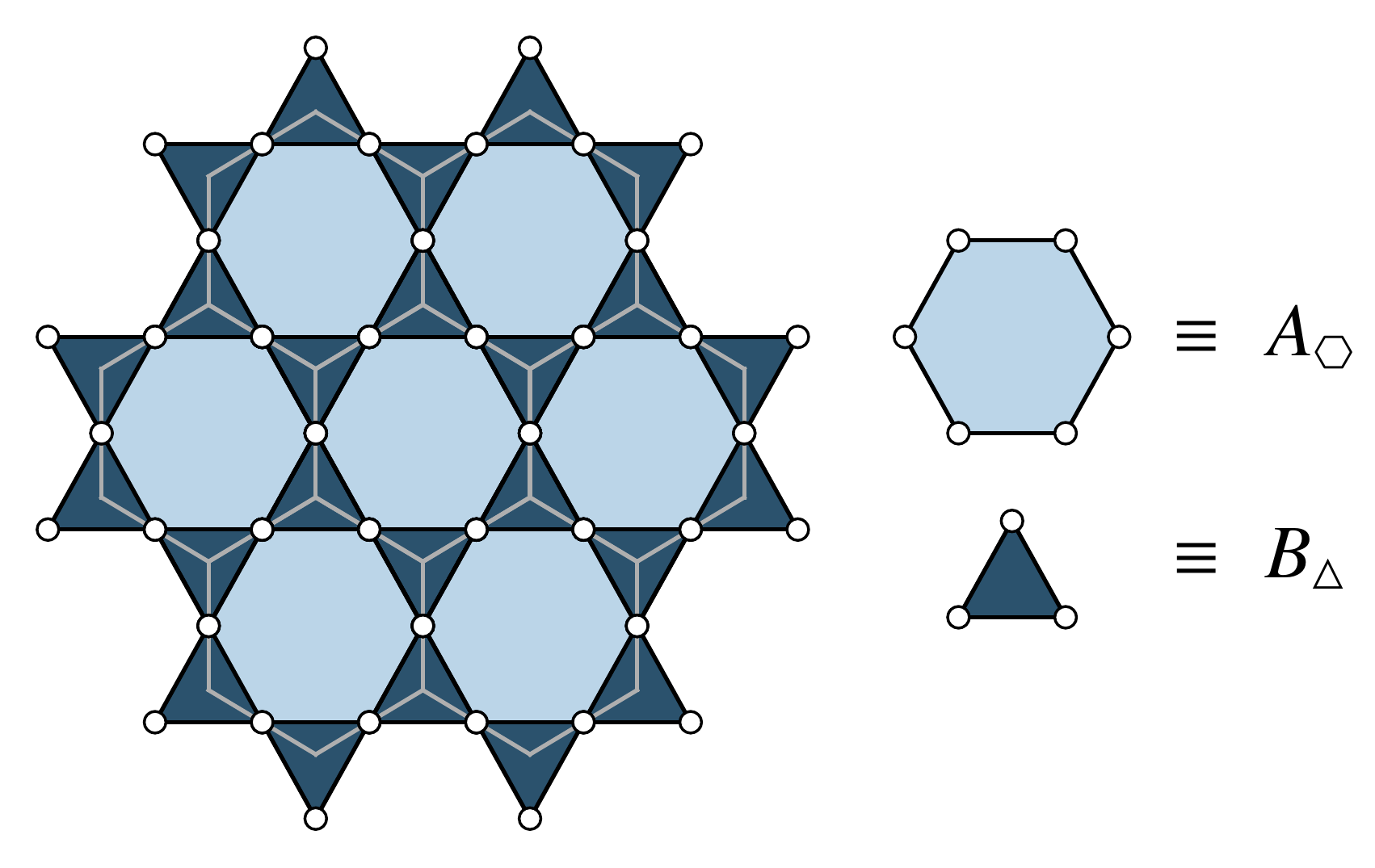}
    \caption{A kagome lattice of spins, depicted by the white
circles, and the corresponding plaquette operators, $A_{\scaleto{\hexagon}{3.5pt}}$ and $B_{\scaleto{\triangle}{3.5pt}}$, which comprise the toric code Hamiltonian~\eqref{eqn:kagome-code}. The centres of the hexagonal plaquettes, $A_{\scaleto{\hexagon}{3.5pt}}$, form a triangular lattice, while the centres of the triangular plaquettes, $B_{\scaleto{\triangle}{3.5pt}}$, form a hexagonal lattice.}
    \label{fig:kagome-code}
\end{figure}

In this section we calculate explicitly the generating functions for nonreversing walks on the triangular and honeycomb lattices. Both lattices appear naturally in the context of frustrated magnetism. However, as a concrete example, consider the toric code defined on the kagome lattice, as in Fig.~\ref{fig:kagome-code}:
\begin{equation}
    H = -\lambda_A \sum_{\hexagon} A_{\scaleto{\hexagon}{3.5pt}} - \lambda_B \sum_{\triangle} B_{\scaleto{\triangle}{3.5pt}}
    \, ,
    \label{eqn:kagome-code}
\end{equation}
where $A_{\scaleto{\hexagon}{3.5pt}} = \prod_{i \in {\scaleto{\hexagon}{3.5pt}}} \sigma_i^x$,  $B_{\scaleto{\triangle}{3.5pt}} = \prod_{i \in {\scaleto{\triangle}{3.5pt}}} \sigma_i^z$, corresponding to the hexagonal ($\hexagon$) and triangular ($\triangle$, both `up' and `down') plaquettes of the lattice, respectively.
$\lambda_{A(B)}>0$ are the two coupling constants of the model.
The operators are all mutually commuting, $[A_{\scaleto{\hexagon}{3.5pt}}, B_{\scaleto{\triangle}{3.5pt}}]=0$, since each hexagonal plaquette shares an even number of spins with any overlapping triangular plaquette.

As in the case of the square lattice, the cases $\lambda_A \ll \lambda_B$ ($\lambda_B \ll \lambda_A)$ can be generated perturbatively in the ground state sector by applying a small magnetic field in the $x$ ($z$) direction to a system with $\lambda_{A(B)}=0$.
We will use the terminology that the lower-energy excitations, generated perturbatively via ring exchange, correspond to the visons.
If the visons reside on the triangular plaquettes, then the spinons, which live on the hexagonal plaquettes, hop on a triangular lattice.
Conversely, in the opposite limiting case, if the visons live on the hexagonal plaquettes, then the spinons hop on a hexagonal lattice.

%
%

\subsection{Triangular lattice}

For the triangular lattice the coordination number $z=6$ and there are thus six possible moves at each step, enumerated by the generating variables $\delta$ and
$\epsilon$:
$\delta$, $\epsilon$, $\epsilon^{-1}$, $\delta^{-1}$, $\epsilon\delta^{-1}$ and $\delta\epsilon^{-1}$. The matrix which governs transitions between these various allowed moves between adjacent sites is
\begin{equation}
    N = x
    \begin{pmatrix}
        \epsilon & \delta & \epsilon^{-1}\delta & 0 &  \delta^{-1} & \epsilon \delta^{-1} \\
        \epsilon & \delta & \epsilon^{-1}\delta & \epsilon^{-1} & 0 & \epsilon \delta^{-1} \\
        \epsilon & \delta & \epsilon^{-1}\delta & \epsilon^{-1} & \delta^{-1} & 0 \\
        0 & \delta & \epsilon^{-1}\delta & \epsilon^{-1} & \delta^{-1} & \epsilon \delta^{-1} \\
        \epsilon & 0 & \epsilon^{-1}\delta & \epsilon^{-1} & \delta^{-1} & \epsilon \delta^{-1} \\
        \epsilon & \delta & 0 & \epsilon^{-1} &  \delta^{-1} & \epsilon \delta^{-1} \\
    \end{pmatrix}
    \, ,
\end{equation}
where the zeros enforce the nonreversing constraint imposed on the lattice walk. The initial condition

\begin{equation}
    N_0 = x\diag(\epsilon, \delta, \epsilon^{-1}\delta, \epsilon^{-1} , \delta^{-1}, \epsilon \delta^{-1})
    \, ,
\end{equation}
represents the unconstrained first step.
Using the general expression~\eqref{eqn:nonreversing-gf-general} presented in the main text, the generating function for nonreversing walks is therefore
\begin{equation}
    \mc{N}(x; \delta, \epsilon) = \frac{1-x^2}{1-x(\epsilon + \epsilon^{-1} + \delta + \delta^{-1} + \epsilon \delta^{-1} + \delta \epsilon^{-1}) + 5x^2}
    \, .
\end{equation}
The expression for $\mc{R}_2(x)$ presented in the main text in~\eqref{eqn:R_2k-vs-N} must be generalised to allow for the two basis vectors to be non-orthonormal, i.e., when $\v{e}_{i}\cdot \v{e}_j = a^2\delta_{ij} + a^2(1-\delta_{ij})\cos\theta$, the expression for $\mc{R}_2$ becomes
\begin{equation}
    \mc{R}_2(x) = a^2 \left\{ \left[ \left( \delta \partial_\delta \right)^2 + 2\cos\theta \left( \delta \partial_\delta \right)\left( \epsilon \partial_\epsilon \right) +  \left( \epsilon \partial_\epsilon \right)^2  \right] \mc{N} \right\} \bigg\rvert_{\delta=\epsilon=1}
    \, .
\end{equation}
By symmetry, $\delta \partial_\delta$ and $\epsilon \partial_\epsilon$ commute when acting on $\mc{N}(x; \delta, \epsilon)$.
Evaluating the appropriate derivatives of $\mc{N}(x; \delta, \epsilon)$ we arrive at
\begin{equation}
    \mc{R}_2(x) = \frac{4x(1+x)}{(1-5x)^2(1-x)}(2-\cos\theta)
    \, .
\end{equation}
This leads to the final expression for the diffusion constant
\begin{align}
    D_6 &= \frac{3}{\pi}\left[ 3\sqrt{5} - 2\ln\left( \frac{3+\sqrt{5}}{3-\sqrt{5}} \right) \right]\\
    &= 2.72968\ldots
\end{align}
%
%

\subsection{Honeycomb lattice}

The case of the honeycomb lattice ($z=3$) is complicated slightly by its two-sublattice structure.
We proceed by constructing two generating functions $\mc{N}_{aa}$ and $\mc{N}_{ab}$, corresponding to walks that begin and end on the same sublattice, and walks that begin and end on complementary sublattices, respectively.
Beginning with $\mc{N}_{aa}$, we divide each walk into segments of length 2. Taking into account the nonreversing constraint, there are six possible transitions for each length-2 segment: $\delta$, $\epsilon$, $\epsilon^{-1}$, $\delta^{-1}$, $\epsilon\delta^{-1}$ and $\delta\epsilon^{-1}$, corresponding to moves on the underlying triangular lattice. At each step following the initial one, \emph{two} of these moves are disallowed by the nonreversing constraint leading to the transition matrix
\begin{equation}
    N = x^2
    \begin{pmatrix}
        \epsilon & \delta & 0 & 0 & \delta^{-1} & \epsilon \delta^{-1} \\
        \epsilon & \delta & \epsilon^{-1}\delta & \epsilon^{-1} & 0 & 0 \\
        0 & 0 & \epsilon^{-1}\delta & \epsilon^{-1} &  \delta^{-1} & \epsilon \delta^{-1} \\
        \epsilon & \delta & \epsilon^{-1}\delta & \epsilon^{-1} & 0 & 0 \\
        0 & 0 & \epsilon^{-1}\delta & \epsilon^{-1} & \delta^{-1} & \epsilon \delta^{-1} \\
        \epsilon & \delta & 0 & 0 & \delta^{-1} & \epsilon \delta^{-1} \\
    \end{pmatrix}
    \, ,
\end{equation}
with the initial condition
\begin{equation}
    N_0 = x^2 \diag(\epsilon, \delta, \epsilon^{-1}\delta, \epsilon^{-1}, \delta^{-1},\epsilon \delta^{-1} )
    \, .
\end{equation}
These matrices lead to the generating function
\begin{multline}
    \mc{N}_{aa}(x; \delta, \epsilon) = \\
    1 + \frac{ x^2\left( \epsilon + \epsilon^{-1} + \delta + \delta^{-1} + \epsilon\delta^{-1} + \delta\epsilon^{-1} \right) - 6x^2 }{ 1 - x^2 \left( \epsilon + \epsilon^{-1} + \delta + \delta^{-1} + \epsilon\delta^{-1} + \delta\epsilon^{-1} - 1 \right) + 4x^4 }
    \, .
\end{multline}

For the generating function $N_{ab}$, we write a walk from $a \to b$ as (i) the first step takes the walker from the $a$ to the $b$ sublattice, and (ii) the walker then performs a walk amongst sites belonging to the $b$ sublattice only. This walk is implemented using the matrix
\begin{equation}
    N = x^2
    \begin{pmatrix}
        \epsilon & \delta & 0 & \epsilon^{-1} & 0 & \epsilon \delta^{-1} \\
        \epsilon & \delta & 0 & \epsilon^{-1} & 0 & \epsilon \delta^{-1} \\
        0 & \delta & \epsilon^{-1}\delta & \epsilon^{-1} &  \delta^{-1} & 0 \\
        0 & \delta & \epsilon^{-1}\delta & \epsilon^{-1} &  \delta^{-1} & 0 \\
        \epsilon & 0 & \epsilon^{-1}\delta & 0 & \delta^{-1} & \epsilon \delta^{-1} \\
        \epsilon & 0 & \epsilon^{-1}\delta & 0 &  \delta^{-1} & \epsilon \delta^{-1} \\
    \end{pmatrix}
    \, ,
    \label{eqn:honeycomb-N}
\end{equation}
and the initial condition
\begin{equation}
    N_0 = x N_1 + x \epsilon^{-1} N_2 + x \delta^{-1} N_3
    \, ,
\end{equation}
which represents the three possible moves in the unconstrained first step, where $N_i = \v{e}_{2i} \otimes \v{e}_{2i}$, and $\v{e}_\mu$ ($\mu=1, \ldots, 6$) are the orthonormal basis vectors with respect to which~\eqref{eqn:honeycomb-N} is expressed. These matrices lead to
\begin{multline}
    \mc{N}_{ab}(x; \delta, \epsilon) = \\
    \frac{ x (1 - x^2) (1 + \delta^{-1} + \epsilon^{-1}) }{ 1 - x^2 \left( \epsilon + \epsilon^{-1} + \delta + \delta^{-1} + \epsilon\delta^{-1} + \delta\epsilon^{-1} - 1 \right) + 4x^4 }
    \, .
\end{multline}

The full generating function is then given by $\mc{N} = \mc{N}_{aa} + \mc{N}_{ab}$.
However, in order to calculate $\mc{R}_2$, one should in principle account for the fact that the $b$ sublattice is translated by one lattice constant with respect to the $a$ sublattice. This detail is only relevant for short times, and hence does not need to be taken into account for the calculation of the diffusion constant, which depends only on the asymptotic behaviour of $\langle \v{r}^2(t) \rangle$.

Combining all of the above results gives us the generating function $\mc{R}_2$:
\begin{multline}
    \mc{R}_2(x) = \frac{2a^2}{\left(1-4 x^2\right)^2 \left(1-x^2\right)} \Big\{ \\
    4x^2(1+2x^2) + x(1+7x^2 + 4x^4) - \\
    \cos\theta \left[ 2x^2(1+2x^2) + 6x^3 \right] \Big\}
    \, ,
\end{multline}
where $a=\sqrt{3}$ and $\theta = \pi/3$. Inserting these values simplifies the expression to
\begin{equation}
    \mc{R}_2(x) = \frac{6x \left(1+2x^2\right)}{(1-2 x)^2 (1-x) (1+2x)}
    \, .
\end{equation}
Hence, we arrive at the expression
\begin{equation}
    D_3 = \frac{3}{2\pi} \left[ 12\sqrt{2}-\ln \left( \frac{3+2\sqrt{2}}{3-2\sqrt{2}} \right) \right] = 3.20977\ldots
    \, .
\end{equation}
%
%

\section{Numerical verification}
\label{sec:numerical-verification}

\begin{figure}[t]
    \centering
    \includegraphics[width=\linewidth]{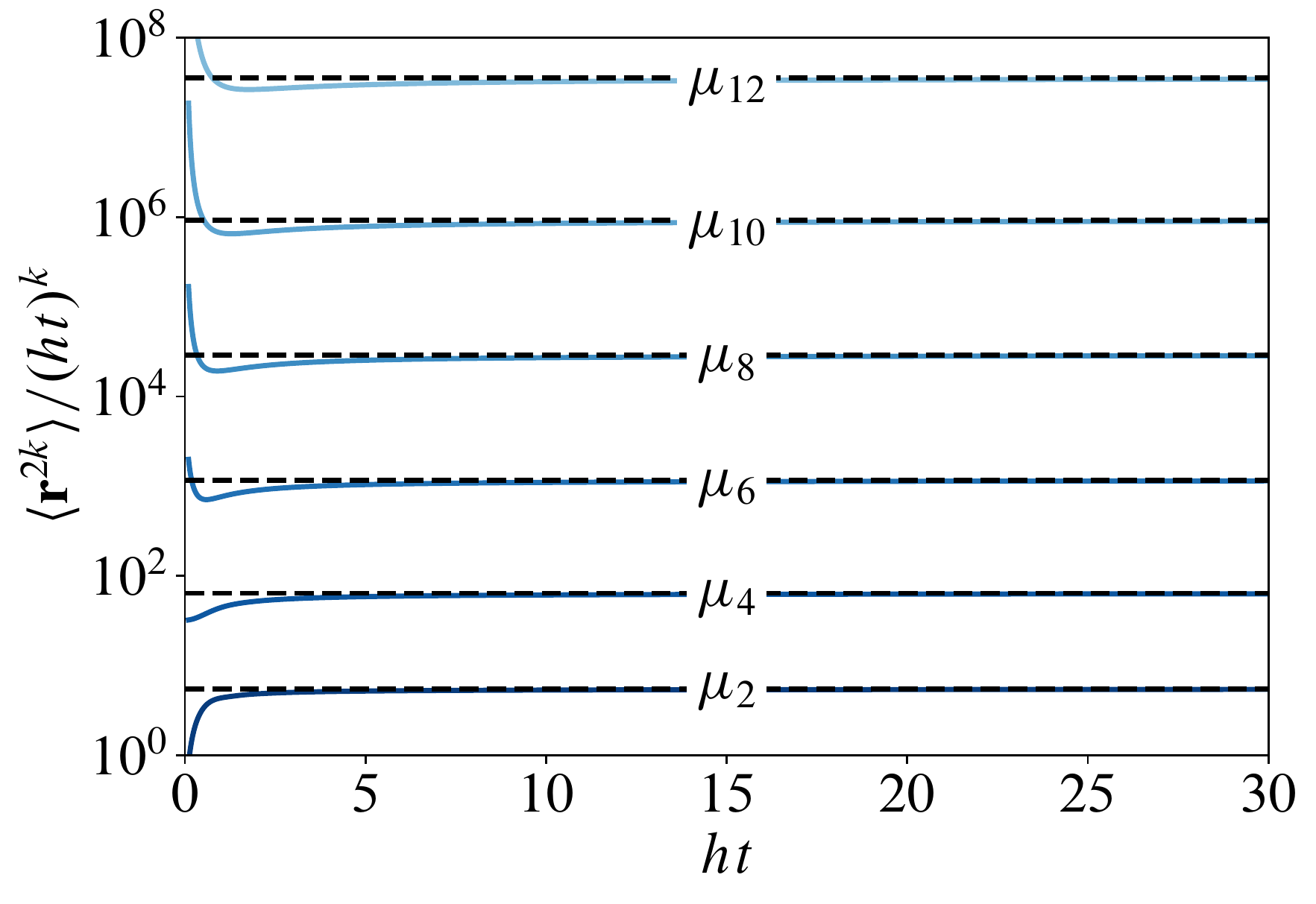}
    \caption{The first six nonzero moments of the density distribution, evaluated numerically (solid lines), exhibiting convergence for large times towards the analytical expression~\eqref{eqn:general-moments} (dashed lines) presented in the main text.}
    \label{fig:numerical-moments}
\end{figure}

In this Appendix we confirm numerically our results for the moments of the spinon density profile on the square lattice via an independent calculation.
We perform this verification by calculating numerically how a single particle spreads with time on the Bethe lattice (at zero temperature), and then mapping sites on the Bethe lattice to sites on the square lattice in order to correctly account for the distance of each site from the origin. Suppose that the particle's wave function has the projection $\psi_\ell(t)$ onto a site at depth $\ell$ on the Bethe lattice. The symmetry of the lattice dictates that the magnitude of this projection is the same for all sites at the same depth. Then,
\begin{equation}
    \langle \v{r}^{2k}(t) \rangle = \sum_{\ell=0}^\infty \sum_{s_\ell=1}^{N_\ell} \v{r}_{s_\ell}^{2k} \abs{\psi_\ell(t)}^2
    \, .
\end{equation}

The mapping between the square and Bethe lattices comes from the term $\sum_{s_\ell=1}^{N_\ell} \v{r}_{s_\ell}^{2k}$,
where each site $s_\ell$ on the Bethe lattice maps to a site with position $\v{r}_{s_\ell}$ in real space (i.e., on the original square lattice).
The number of sites at depth $\ell$ is $N_\ell = 4\cdot 3^{\ell-1}$ for $\ell > 0$.
This sum can be evaluated by taking appropriate derivatives of the function $\mc{R}_{2k}(x)$, defined by~\eqref{eqn:R_2k-vs-N}:
\begin{equation}
    \sum_{s_\ell=1}^{N_\ell} \v{r}_{s_\ell}^{2k} = \frac{1}{\ell!} [\partial_x^\ell \mc{R}_{2k}](0)
    \, .
    \label{eqn:average-distance}
\end{equation}
For the special case $k=1$,
\begin{equation}
    \frac{1}{N_\ell}\sum_{s_\ell=1}^{N_\ell} \v{r}_{s_\ell}^{2} = 2\ell - \tfrac32 (1 - 3^{-\ell}) \sim 2\ell
    \, ,
\end{equation}
in agreement with the expression presented in Ref.~\onlinecite{Kanasz-Nagy2017}, obtained by different means. For general $k$, one may show from the general expression~\eqref{eqn:average-distance} that asymptotically
\begin{equation}
    \frac{1}{N_\ell}\sum_{s_\ell=1}^{N_\ell} \v{r}_{s_\ell}^{2k} \sim k! (2\ell)^k
\end{equation}
for sufficiently large depths $\ell$. Hence, the moments may be calculated for sufficiently large times using the expression $\langle \v{r}^{2k}(t) \rangle \!\sim\! k! \sum_\ell N_\ell (2\ell)^k \abs{\psi_\ell(t)}^2 $. This expression is used to evaluate the moments numerically in Fig.~\ref{fig:numerical-moments}, showing convergence towards our exact expression.
%
%

\section{Return probability}
\label{sec:return-probability}

The probability that the spinon returns to its initial site (0) at high temperature (i.e., the survival probability) is found by evaluating all closed, self-retracing walks on the original lattice $\mc{L}_z$. Such walks are enumerated by the generating functions $C_0(x)$ (corresponding to \emph{closed}, nonreversing base paths), and $T(x)$, $S(x)$ (corresponding to self-retracing excursions).

In the case of the square lattice, the double contour integral, \eqref{eqn:general-probability-distribution}, may be rewritten in the form of a surface integral over a square in the $\theta_1$-$\theta_2$ plane:
\begin{multline}
    P_0(t) = -\sum_{\alpha_1,\alpha_2}g_{\alpha_1\alpha_2} \int_0^\pi \int_0^\pi \prod_{j=1}^2\frac{\mathrm{d}\theta_j}{2\pi}\, e^{2\sqrt{3}iht(\cos\theta_1 - \cos\theta_2)} \\
    C_0\left[\tfrac13 e^{i(\alpha_1 \theta_1 + \alpha_2 \theta_2)} \right]\prod_{j=1}^2\frac{\sin\theta_j}{(\tfrac43 - \cos^2\theta_j)}(\cos\theta_j - 2e^{i\alpha_j\theta_j})
    \, ,
    \label{eqn:return-probability-shrunk}
\end{multline}
\begin{figure}[t]
    \centering
    \includegraphics[width=\linewidth]{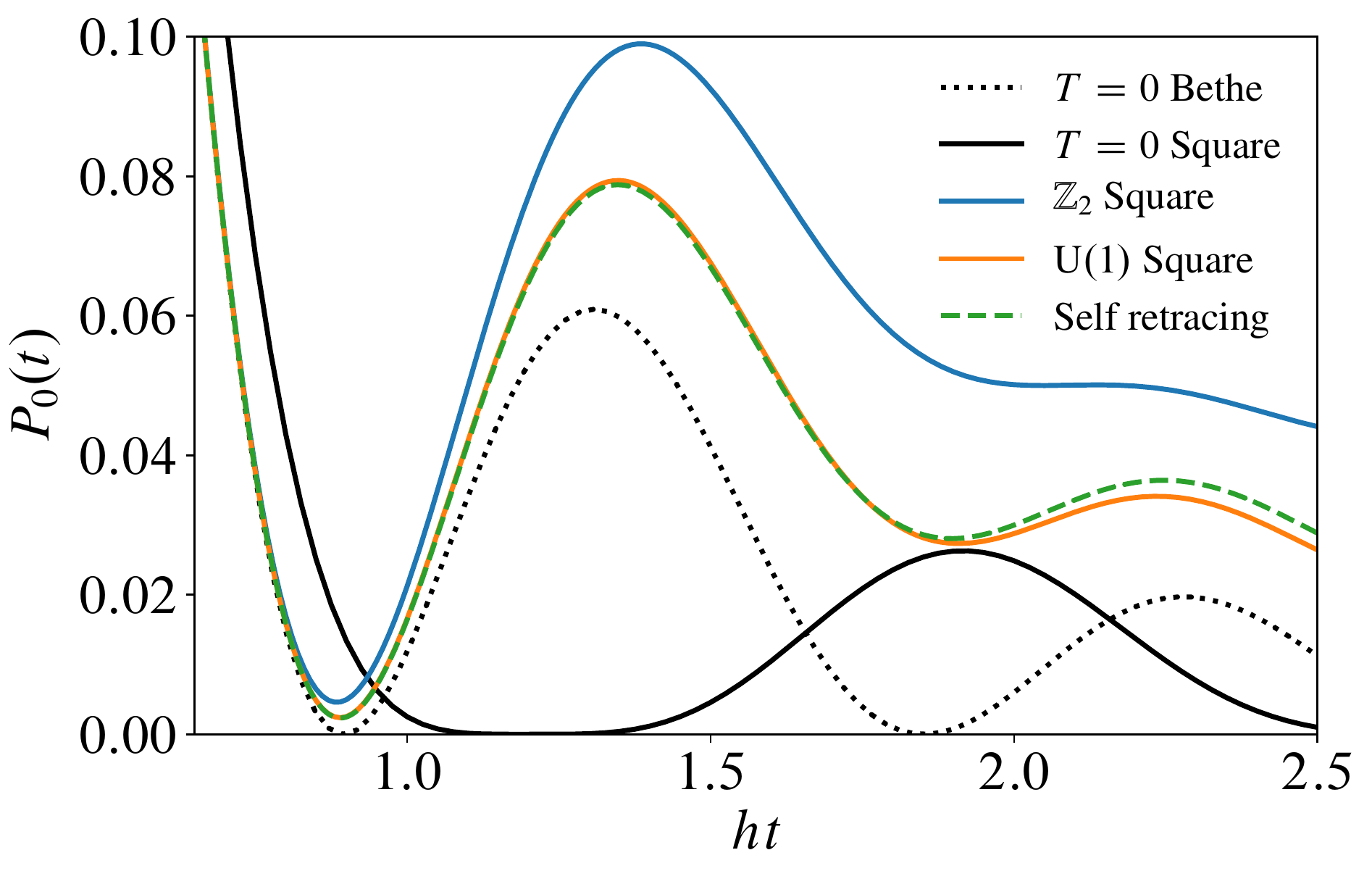}
    \caption{Short-time behaviour of the spinon's return probability, $P_0(t)$.
    The calculation is performed on a $49 \times 49$ square lattice and the data are averaged over $\num{25000}$ histories.
    The full self-retracing result, \eqref{eqn:general-probability-distribution}, differs qualitatively from the $T=0$ return probability on the Bethe lattice in that the latter periodically exhibits perfect destructive interference and decays more rapidly with time.}
    \label{fig:return-probability}
\end{figure}
where the symmetric matrix
\begin{equation}
    g_{\alpha_1 \alpha_2} = e^{i\pi (\alpha_1 - \alpha_2)/2} =
    \begin{pmatrix}
        \phantom{-}1 & -1 \\
        -1 & \phantom{-}1
    \end{pmatrix}_{\alpha_1 \alpha_2}
    \, ,
\end{equation}
ensures that the probability $P_0(t)$ remains real for all time.
We remind the reader that
\begin{align}
    C_0(x) &= \frac{2}{\pi} \left( \frac{1-x^2}{1+3x^2} \right) K \left( \frac{4x}{1+3x^2} \right) \\
    &= \includegraphics[height=4pt, valign=c]{path_0.pdf} + (8 \, \includegraphics[height=8.5pt, valign=c]{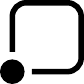})x^4 + (8 \, \includegraphics[height=8.5pt, valign=c]{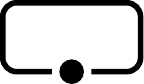} + 16 \, \includegraphics[height=8.5pt, valign=c]{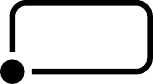} + 16 \, \includegraphics[height=8.5pt, valign=c]{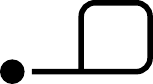})x^6 + \ldots
    \, ,
\end{align}
where $K(x)$ is the complete elliptic integral of the first kind. Asymptotically, as $ht \to \infty$, within the Bethe lattice approximation, the return probability equals
\begin{equation}
    P_0(t) = \frac{c_1}{t} + \frac{c_2}{t^2} \cos(4\sqrt{3}ht) + \ldots
    \, ,
\end{equation}
where the $c_i$ are $O(1)$ constants. The first term is consonant with the asymptotic diffusion of the density profile, while the latter term corresponds to transient coherent oscillations.

The exact expression~\eqref{eqn:return-probability-shrunk} is plotted in Fig.~\ref{fig:return-probability} and compared with the corresponding numerics for the continuous and $\pi$-flux models.
The form of the return probability differs drastically from the zero temperature result corresponding to ballistic propagation of the spinon, where $P_0(t) \sim \cos^4(2ht - \pi/4)/t^2$ for $ht \gg 1$, in which the survival probability decays more rapidly ($\sim\! t^{-2}$) and the coherent oscillations persist indefinitely.
The discrepancy between the Bethe lattice result and the high-temperature limit of the $\pi$-flux model at these short times is attributed to the neglect of loop diagrams of the form shown in Fig.~\ref{fig:paths:b}.
%
%

\bibliographystyle{aipnum4-1}
\bibliography{references}

\begin{thebibliography}{68}
\bibitem [{Note11()}]{regularisation}%
  \BibitemOpen
  \bibinfo {note} {The expression for the length $\xi _c$ in the continuous
  case must be regularised at large temperatures. We use $\xi _c^4 = 1 + 4\beta
  ^2\Delta_v^2$.}\BibitemShut {Stop}%
\end{thebibliography}

\end{document}